\documentclass[a4paper,fleqn]{cas-sc}

\usepackage{amsmath, amssymb, amsfonts, dcolumn, graphicx, float, hyperref, xcolor, placeins, multirow, hhline, threeparttable}
\usepackage[authoryear]{natbib} 

\def\tsc#1{\csdef{#1}{\textsc{\lowercase{#1}}\xspace}}
\tsc{WGM}
\tsc{QE}

\begin{document}
\let\WriteBookmarks\relax
\def\floatpagepagefraction{1}
\def\textpagefraction{.001}

\shorttitle{Photometric observation of a QSME in 2021}
\shortauthors{C.W. So, et. al.}
\title[mode = title]{The photometric observation of the quasi-simultaneous mutual eclipse and occultation between Europa and Ganymede on 22 August 2021}

\author[1]{Chu Wing So}
\ead{cwso@lcsd.gov.hk}
\cormark[1]
            
\author[2]{Godfrey Ho Ching Luk}
\ead{1155159752@link.cuhk.edu.hk}

\author[2]{Giann On Ching Chung}
\ead{1155143045@link.cuhk.edu.hk}
            
\author[2]{Po Kin Leung}
\ead{pkleung@cuhk.edu.hk}

\author[3]{Kenneith Ho Keung Hui}
\ead{kenneith@hokoon.edu.hk}

\author[1]{Jack Lap Chung Cheung}
\ead{jackcheung1996@gmail.com}

\author[1]{Ka Wo Chan}
\ead{markkawochan@gmail.com}

\author[4]{Edwin Lok Hei Yuen}
\ead{lhyuene@connect.hku.hk}

\author[1]{Lawrence Wai Kwan Lee}
\ead{wklee@lcsd.gov.hk}

\author[5]{Patrick Kai Ip Lau}
\ead{pkiplau@lcsd.gov.hk}

\author[1]{Gloria Wing Shan Cheung}
\ead{gwscheung@lcsd.gov.hk}

\author[1]{Prince Chun Lam Chan}
\ead{chlchan@lcsd.gov.hk}

\author[4]{Jason Chun Shing Pun}
\ead{jcspun@hku.hk}

\affiliation[1]{organization={Hong Kong Space Museum},
            addressline={10 Salisbury Road, Tsim Sha Tsui, Kowloon}, 
            city={Hong Kong},
            country={China}}
            
\affiliation[2]{organization={Department of Physics, The Chinese University of Hong Kong},
            addressline={Sha Tin, New Territories}, 
            city={Hong Kong},
            country={China}}    
            
\affiliation[3]{organization={Ho Koon Nature Education cum Astronomical Centre (Sponsored by Sik Sik Yuen)},
            addressline={101 Route Twisk, Tsuen Wan, New Territories}, 
            city={Hong Kong},
            country={China}}   
            
\affiliation[4]{organization={Department of Physics, The University of Hong Kong},
            addressline={Pok Fu Lam}, 
            city={Hong Kong},
            country={China}}  
            
\affiliation[e]{organization={Hong Kong Science Museum},
            addressline={2 Science Museum Road, Tsim Sha Tsui East, Kowloon}, 
            city={Hong Kong},
            country={China}}    
        
\begin{abstract}
Mutual events (MEs) are eclipses and occultations among planetary natural satellites. Most of the time, eclipses and occultations occur separately. However, the same satellite pair will exhibit an eclipse and an occultation quasi-simultaneously under particular orbital configurations. This kind of rare event is termed as a quasi-simultaneous mutual event (``QSME'').
During the 2021 campaign of mutual events of jovian satellites, we observed a QSME between Europa and Ganymede. The present study aims to describe and study the event in detail.
We observed the QSME with a CCD camera attached to a 300-mm telescope at the Hong Kong Space Museum Sai Kung iObservatory. We obtained the combined flux of Europa and Ganymede from aperture photometry. A geometric model was developed to explain the light curve observed. Our results are compared with theoretical predictions (``O-C'').
We found that our simple geometric model can explain the QSME fairly accurately, and the QSME light curve is a superposition of the light curves of an eclipse and an occultation. Notably, the observed flux drops are within 2.6\% of the theoretical predictions. The size of the event central time O-C's ranges from $-14.4$ to 43.2~s. Both O-C's of flux drop and timing are comparable to other studies adopting more complicated models.
Given the event rarity, model simplicity and accuracy, we encourage more observations and analysis on QSMEs to improve Solar System ephemerides.
\end{abstract}


\maketitle

\section{Introduction} 
\label{sec:introudction}

Mutual events (MEs or mutual phenomena) are eclipses and occultations among
planetary satellites. For the jovian, saturnian and uranian systems, a series
of MEs happen about every 6, 14 and 42~years respectively~\citep{arlot(2019)}
when the Sun crosses the planetary equatorial plane (mutual eclipse) or when
Earth crosses such a plane (mutual occultation). ME observations have been
extended to binary asteroids~\citep{descamps(2008),berthier(2020)}.

Numerous observations have been made on MEs whenever the occasion arose since
the 1970s (see~\citet{arlot(2019)} for a recent review). Much information about
the satellite systems can be obtained from ME observations. For example, since
the reduction of the satellites' flux during ME depends on how sunlight is
reflected or scattered from the satellites, results from ME photometric
observations place constraints on the eclipsed/occulted satellite's albedo,
surface hot spots associated with
volcanism~\citep{fujii(2014),descamps(1992a),goguen(1988)}, limb darkening and
atmosphere existence~\citep{Aksnes(1974)}, etc. 

The positions of the satellites also determine the details of flux reduction.
Extracting astrometric information from the ME's photometric data has been one
of the direct ways to construct satellite
ephemerides~\citep{arlot(2019),saquet(2016),lainey(2004b)}. On the other hand,
the comparisons between observations and theoretical computations (known as
``O-C'') facilitate the improvement of existing satellite ephemerides and the
theories of satellite
motion~\citep{emelyanov(2020b),arlot(2019),saquet(2018),saquet(2016),arlot(2014),emelyanov(2013),arlot(2009),emelyanov(2002)}. 

ME astrometry can now achieve an accuracy close to that of HST
observations~\citep{arlot(2019)} using sensitive semiconductor sensors such as
CCD cameras. By analysing highly accurate and long-term data, one can reveal
the satellite's interior structure and
dynamics~\citep{arlot(2019),arlot(2014),emelyanov(2011),arlot(2009),vienne(2008),aksnes(2001)},
confirm the depth of oceans~\citep{noyelles(2004)} and the planet--satellite
tidal dissipation~\citep{lainey(2009)}. 

\begin{figure}[ht] 
\centering
\includegraphics[width=9 cm]{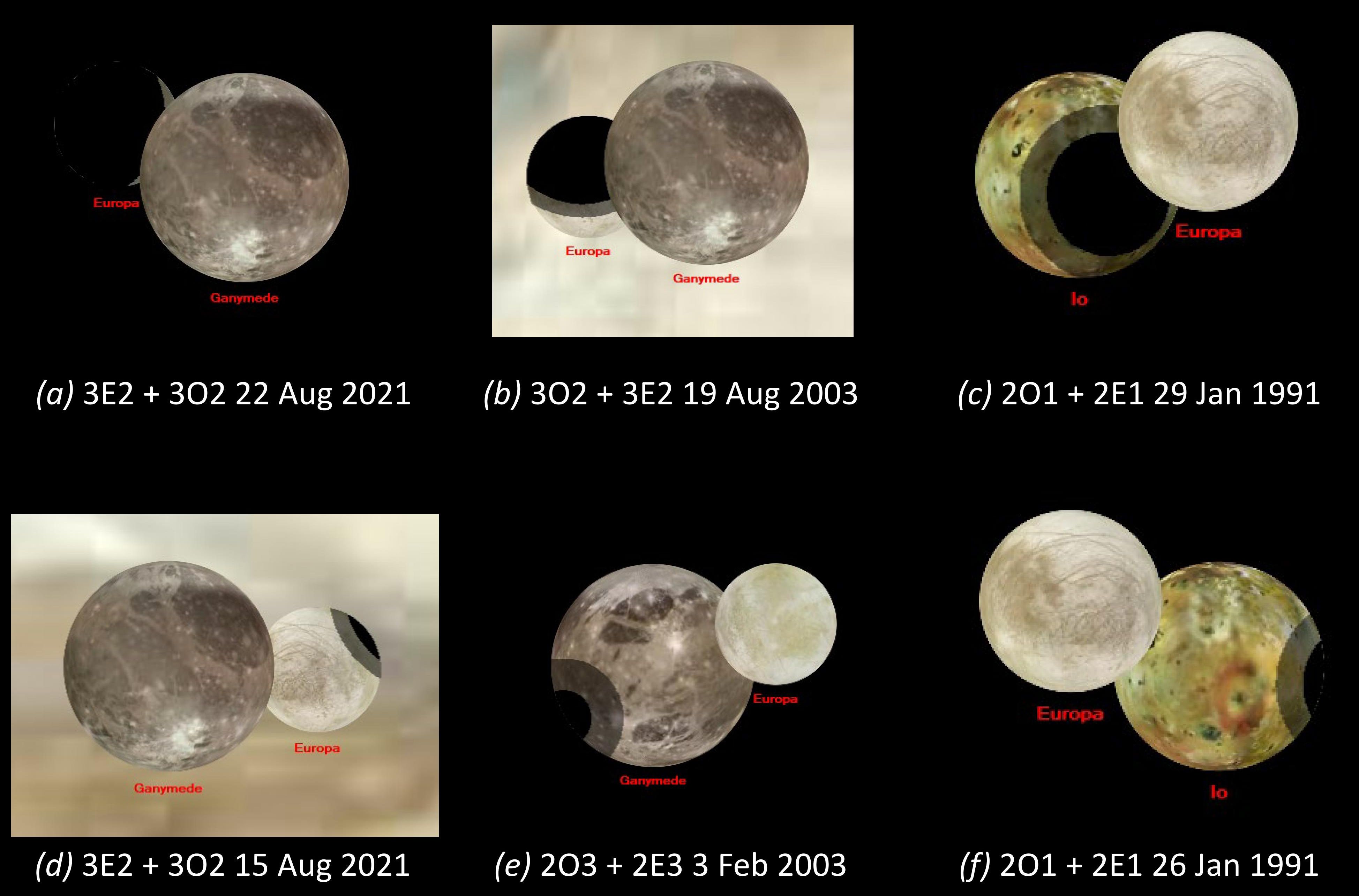}
\caption{Simulated examples of QSMEs. 
\textit{(a)}--\textit{(c)} are events with shadows overlapping with the active satellites, while \textit{(d)}--\textit{(f)} are events without shadows overlapping. 
\textit{(a)} is the event in the present study.
\textit{(c)} is the event described in~\citet{vasundhara(1994)}. 
\textit{(b)} \& \textit{(d)} happened in-front-of Jupiter. 
Different events are not drawn in the same scale. The simulations are created with WinJUPOS (\url{http://www.grischa-hahn.homepage.t-online.de/}).\label{fig:QSME_examples}}
\end{figure}

Jupiter's Galilean satellites (J1: Io; J2: Europa; J3: Ganymede; J4: Callisto)
are the most studied dynamical systems. Their MEs have received much attention
because they are relatively easier to observe and occur more frequently. In
addition, due to the absence of atmosphere on Galilean satellites, the accuracy
of ME astrometry can exceed the diffraction limit of a
telescope~\citep{vienne(2008)} or has a 1$\sigma$ accuracy of about
$0.025''$~\citep{lainey(2009)}. However, almost all of the previous studies
of Galilean MEs limited the scope to individual eclipses or occultations only.
Indeed, if MEs occur near the time of jovian opposition, mutual eclipses and
occultations may happen almost simultaneously as the shadow of the active
satellite is very close to (or even overlaps with) the active satellite itself
from the Earth's perspective. Fig.~\ref{fig:QSME_examples} shows simulated
examples. It is also possible that the shadow of another satellite (other than
the occulting satellite) projects on the passive satellite which is being
occulted (see Sect.~\ref{sec:conclusions}). 
In both cases, an occultation/eclipse will start before the end of an eclipse/occultation on the passive satellite. 
We name this kind of special event a ``quasi-simultaneous mutual event'' or
``QSME''. A QSME's duration is typically longer than that of an ordinary ME
(which takes several minutes to tens of minutes
only~\citep{emelyanov(2020b),arlot(2019)}) and may last for more than an hour.

In this paper, we adopt the standard designating practice: the code 3E2 means J3 eclipses J2; 3O2 means J3 occults J2, etc. 

This study does not focus on ``double eclipse'' (i.e.,~two eclipses happen
quasi-simultaneously,  see~\citet{emelyanov(2022)} for a double eclipse example
occurred on 19 April 2021) or ``double occultation'' (i.e.,~two occultations
happen quasi-simultaneously). Besides, we will not discuss close pairs of
separated MEs in which an eclipse/occultation happens immediately after/before
an eclipse/occultation (e.g.,~see~\citealp{aksnes(1976)},~\citealp{berthier(2020)}). Lastly, we skip the
discussion on the ``simultaneous mutual event'' (``SME''), which is possible in
theory but is probably extremely rare as it requires a pair of ME to occur
\textit{exactly} simultaneously.

QSMEs have been under-researched. To look for a QSME, observers have to go
through the prediction lists and check the time intervals of individual MEs
one-by-one to see if another ME starts before the end of an ME on the passive
satellite. Many other authors and observers were aware of QSMEs, but little
attention was paid to detailed examination. For example,~\citet{price(2000)}
introduced the QSMEs on 18 February 1932 and 19 August 2003 (simulated in
Fig.~\ref{fig:QSME_examples}\textit{(b)}). Some researchers skipped the analysis or
left the analysis for ``future works'' even though several QSMEs in 2003, 2009
and 2014--15 were observed or listed~\citep{pauwels(2005)}. Without going into
further details, other works published the QSME light curves with two minima
for each when QSMEs took place on 26 January 1991
(Fig.~\ref{fig:QSME_examples}\textit{(f)})~\citep{mallama(1992),emelyanov(2020c)}
and 3 February 2003 (Fig.~\ref{fig:QSME_examples}\textit{(e)})~\citep{arlot(2009)}.
Amateur
astronomers\footnote{\url{https://skyandtelescope.org/astronomy-news/mutual-event-season-heats-up-at-jupiter/}.} also observed the QSMEs in 2009 and
2015.  

\citet{noyelles(2003)} analysed a QSME between saturnian satellites Enceladus
and Tethys on 14 September 1995. However, the existence of two light curve
minima was doubtful as the data signal-to-noise ratio was low.

As far as we know, the only work that attempted to analyse jovian QSME was done
by~\citet{vasundhara(1994)}. The author analysed a QSME of 2O1 then 2E1 on 29
January 1991 (simulated in Fig.~\ref{fig:QSME_examples}\textit{(c)}) and described
it as an ``overlapping'' or a ``composite'' event. The light curve was fitted
with theoretical models that help derive the relative astrometric positions of
the satellites. However, we cannot repeat the analysis due to the lack of
published data and sufficient details of the model (but see Sect.~\ref{sec:fit1991}
where we analysed the same event from different observations using our model).

Here we present a QSME observed with a 2-megapixel CCD camera attached to the prime focus of a 300-mm reflector in Sai Kung, Hong Kong on 22 August 2021. We report that the QSME photometric light curve is a kind of combination of eclipse and occultation effects. To characterise the QSME, a new model is developed. The observed light curve is fitted with the model and an O-C analysis is conducted.

This paper is organised as follows. The details of the observations, data reduction and model are presented in Sect.~\ref{sec:observations} to Sect.~\ref{sec:model} respectively.  Sect.~\ref{sec:results} presents the results. Finally, we draw conclusions and initiate discussions in Sect.~\ref{sec:conclusions}.

\section{Observations}
\label{sec:observations}

The Natural Satellites Ephemeride Server
\textit{MULTI-SAT}~\citep{emelyanov(2008b)} predicted 242 Galilean MEs globally
for the PHEMU21 observation campaign in 2021~\citep{arlot(2020)}. Considering
the Jupiter--Sun angular separation, only 192 events were observable from 3
March 2021 to 16 November 2021. See~\citet{emelyanov(2022)} for the report on
the astrometric results of the campaign. QSMEs were rare: there was only one
predicted occurrence (3E2 $+$  3O2, simulated in  Fig.~\ref{fig:QSME_examples}\textit{(a)}) between 13:59 and 15:26 on 22 August (UTC
time, same hereafter). \textit{Occult}
software\footnote{\url{http://www.lunar-occultations.com/iota/occult4.htm}.}
predicted an additional QSME (3O2 $+$  3E2, simulated in  Fig.~\ref{fig:QSME_examples}\textit{(d)}) between 16:24 and 17:46 on 15 August. Neither events were observed in~\citet{emelyanov(2022)}.

We observed the one on 22 August but lost the one on 15 August due to bad weather conditions. The 15 August occurrence was unfavourable for photometric observation because the satellites were in transit across Jupiter during the entire QSME. The jovian disc's brightness overwhelmed the satellite's light in that case. 

Table~\ref{tab:prediction} tabulates the \textit{MULTI-SAT}'s predictions based
on~\citet{lainey(2009)}'s theory for the 22 August event. 

\begin{table}[ht]
\centering
\caption{\textit{MULTI-SAT}'s predictions of the QSME on 22 August 2021. 
\label{tab:prediction}}
\centering
\begin{tabular}{rrr}
\hline
\textbf{ } & \textbf{3E2} & \textbf{3O2} \\
\hline
Begin time, $t_{begin}$ (hour after 13:00) & 0.984 & 1.565 \\ 
Central time, $t_{central}$ (hour after 13:00)  & 1.513 & 2.007 \\
End time, $t_{end}$ (hour after 13:00)     & 2.044 & 2.448 \\ 
Impact parameter, $x$ (arcsec)        & 0.0245   & 0.4053\\
Flux drop ratio, $\Delta I$           & 0.6530   & 0.6584\\
\hline
\end{tabular}
\end{table}

The observation was conducted at the Hong Kong Space Museum Sai Kung
iObservatory (IAU observatory code: D19; telescope position: $22^{\circ}24'29.3''$N, $114^{\circ}19'22.6''$E, 92.5~m above sea level), which is located in Pak Tam within the
Sai Kung West Country Park, New Territories, Hong Kong.
See~\citet{so(2019),so(2014)} and \citet{pun(2013)} for the site's quality, i.e.~light
pollution conditions. We attempted to observe the event independently at the Ho
Koon Nature Education cum Astronomical Centre in Tsuen Wan (about 22.3~km west
of iObservatory), but it was cloudy at that location. Observing steps and
procedures followed~\citet{arlot(2020),arlot(2019)}. 

\begin{figure}[ht] 
\centering
\includegraphics[width=8 cm]{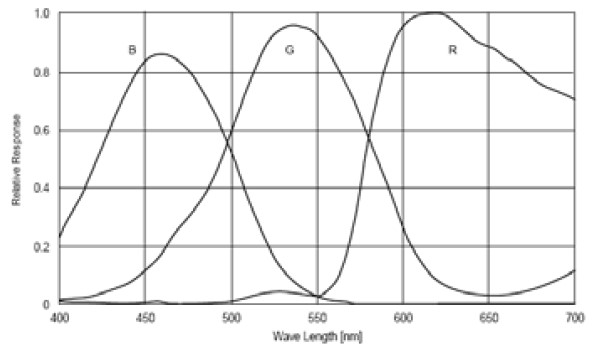}
\caption{Spectral responses of the SKYnyx2-2C camera. Obtained from \url{https://www.lumenera.com/media/wysiwyg/resources/documents/datasheets/astro/skynyx2-2-datasheet.pdf} \label{fig:camera_curve}}
\end{figure}

We used the Takahashi Mewlon-300 Dall--Kirkham Cassegrain telescope (primary mirror of 300~mm diameter, focal length of 2960~mm) on the Takahashi EM-500 Temma II German equatorial mount (tracked at the sidereal rate of 15.041'' s$^{-1}$), equipped with the Lumenera SKYnyx2-2C with a colour CCD chip at the prime focus. The chip has a matrix of 1616$\times$1216 pixels of 4.4$\times$4.4~µm. The setup corresponds to an effective field-of-view (FOV) of 8.4'$\times$6.0'. The chip's spectral responses to different colours are presented in Fig.~\ref{fig:camera_curve}. No additional filter was applied. No track guiding was adopted because the polar alignment was performed accurately. The camera does not have a cooling feature.

We imaged Jupiter, J2 and J3 (both involved in the QSME) and J1 (not involved in the QSME but used as the photometric reference) within the same frame in Flexible Image Transport System (FITS) format. The images were slightly de-focused to avoid saturation while maintaining a reasonable target signal-to-noise ratio. While the telescope's optical axis was slightly misaligned, leading to the deformation of satellite images, the photometric accuracy was unaffected because the stellar aperture included all light from both satellites (see below). Continuous unbinned exposures, each at 1.5~s, were made from 13:55 (around 5~min before the QSME) to 15:56 (around 30~min after the QSME) to measure the satellites' brightness outside the QSME. A total of 3138 sequential images were analysed in this study. We used \textit{MaxIm DL 6} software\footnote{\url{https://diffractionlimited.com/product/maxim-dl/}.} for image acquisition. 

To get accurate timing, the laptop's clock was synchronised with a GPS antenna cum the NTP server Microsemi SyncServer S80, which received time signals directly from the Global Navigation Satellite System (GNSS) satellites. Compared with the method using online time synchronisation services, our method avoided time delay when the time signal passed through the internet. Good GNSS satellite signals with 13 satellite views were received throughout the observation. After getting the shutter latency (0.33~s, measured prior to the observations) \textit{MaxIm DL} timestamped each image to one-hundredth of a second precision. The timestamp was considered as the mid-time of each exposure.

\section{Data reduction}
\label{sec:data_reduction}

\begin{figure}[ht] 
\centering
\includegraphics[width=9 cm]{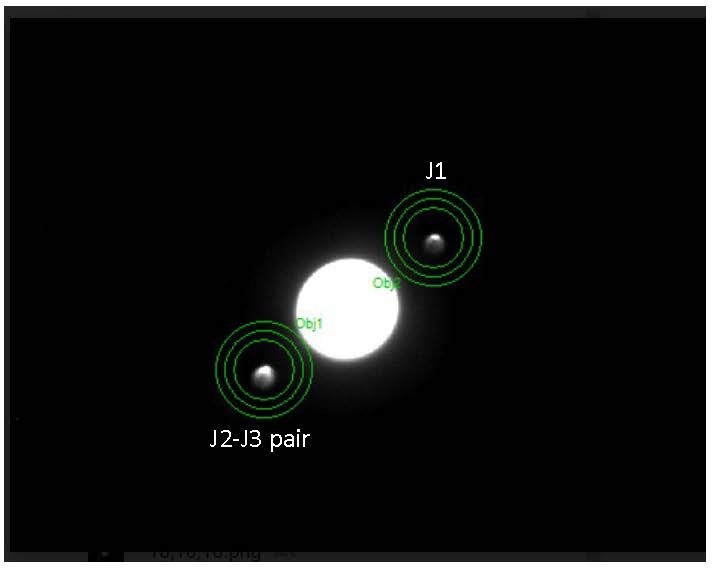}
\caption{Setting of aperture photometry: on each target, the innermost circle shows the aperture radius; the region between the outermost and the middle circles is the sky annulus. The saturated disc at the centre is Jupiter. \label{fig:aperture}}
\end{figure}

We performed aperture photometry separately on the J2--J3 pair and J1 on each image with \textit{MaxIm DL 6}. While it was possible to record the light variations of the eclipsed satellite during an ordinary mutual eclipse in previous studies, the total combined flux from J2 (the eclipsed and occulted satellite) and J3 (the eclipsing and occulting satellite) rather than their individual fluxes was measured because they were indistinguishable in the image most of the time in our case. After some testing, we chose 40 pixels as the optimal aperture radius. Note that this aperture size is relatively large compared to the usual stellar photometry. This aperture is large enough to contain the light from the J2--J3 pair even 30~min after the QSME. Another reason for using a larger-than-usual aperture is that the images were de-focused and exhibited a wider point spread function, as mentioned previously. 
The radius of the sky annulus and its width are 50 and 10~pixels respectively.  Fig.~\ref{fig:aperture} shows the setting overlaid on a sample image. We also conducted photometry independently with the \textit{phot} task under the \textit{NOAO.DIGIPHOT.APPHOT} package in the Image Reduction and Analysis Facility (IRAF) written at the National Optical Astronomy Observatories (NOAO). Similar results were obtained.

To construct the light curve, for each image we calculated the flux ratio $E =f_{\mathrm{event}}/f_{\mathrm{reference}}$, where $f_{\mathrm{event}}$ and $f_{\mathrm{reference}}$ are the fluxes obtained from the J2-J3 event pair and the J1 reference respectively. By dividing, atmospheric influence will be eliminated because the satellites were close to each other in the sky and experienced the (assumed) same atmospheric influence. The bias and dark corrections were incorporated in our data reduction. 

\begin{figure*}[ht] 
\centering
\includegraphics[width=16 cm]{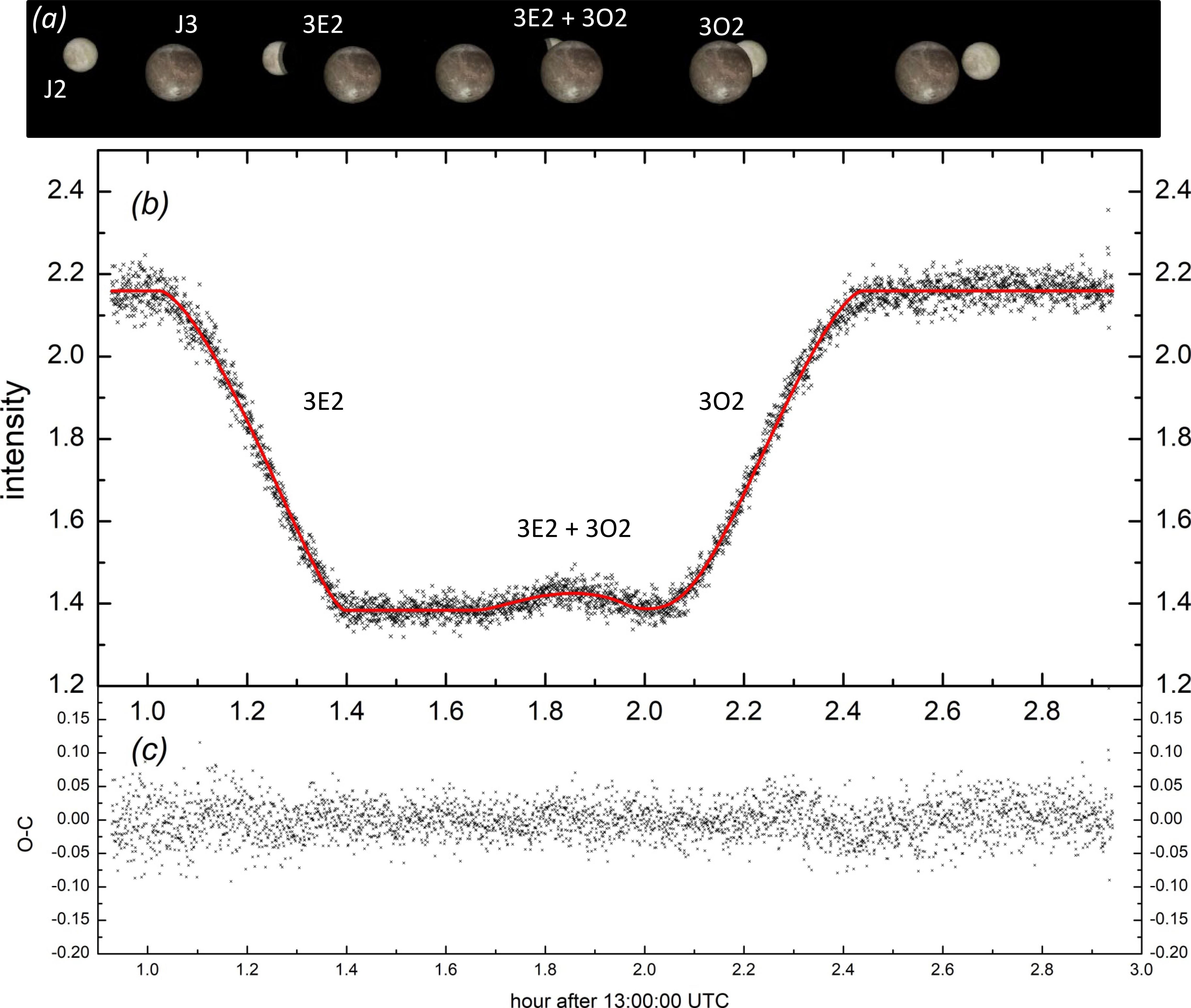}
\caption{The 3E2 $+$  3O2 event on 22 August 2021.
\textit{(a)}: sequential simulations created with WinJUPOS (\url{http://www.grischa-hahn.homepage.t-online.de/});
\textit{(b)}: observed data points and the fitted light curve in red; 
\textit{(c)}: O-C.  
\label{fig:light_curve}}
\end{figure*}

Fig.~\ref{fig:light_curve}\textit{(b)} presents the light curve with very good
data quality. A significant flux drop is evident. For ease of understanding, we
added the event simulation, created based
on~\citet{lainey(2006),lainey(2004a),lainey(2004b)}, in
Fig.~\ref{fig:light_curve}\textit{(a)}. When comparing the simulation with the
observations, one can see that there is a gradual light drop attributed to the
3E2 near the beginning. A part of the observed drop near the middle combines an
eclipse rise and an occultation drop. The combination produces the asymmetric
light curve (ordinary MEs have near symmetric light curves, but
see~\citet{vasundhara(2017)}). There is a ``bump'' with a small (2\%--3\%)
brightness increase peaked at 1.86~h after 13:00 (or near 14:52). The bump is
attributed to the ``short-lived'' period when J2 was leaving J3's shadow and
was soon partially blocked by J3 (Fig.~\ref{fig:light_curve}\textit{(a)}).
Remarkably, the light curve exhibits two minima in which the second minimum
(attributed to 3O2) is slightly brighter than the first one (attributed to 3E2)
since the occultation was not a total occultation while the eclipse was a total
eclipse. Near the end of QSME, the overall brightness returned gradually after
3O2 and levelled off.

A geometrical model can be used to describe the shape of the light curve. The model is the theme of Sect.~\ref{sec:model}. 

\section{QSME model}
\label{sec:model}

As mentioned in Sect.~\ref{sec:introudction}, few researchers have formulated QSME geometric models. We attempt to develop our model to obtain astrometric data from the light curve.  

First, we express the measured flux $E_{i}$ at the $i$th
observation made at time $t_{i}$ as~\citep{emelyanov(2020b)}:
\begin{equation}
 E_{i}= K \cdot S(t_{i}), \text{ for } i=1,2, \ldots,n
\label{eqt:measure_model}
\end{equation}
 where $S(t_{i})$ is the normalised theoretical flux of the satellite(s) at time $t_{i}$, $n$ is the total number of observations on the same event and $K$ is the proportionality constant relating observation and theory. The key of the model is to describe $S$ in the form of:
\begin{equation}
 S(t_{i})=\dfrac{p_{\mathrm{a}} \pi r_{\mathrm{o}}^{2}+p_{\mathrm{p}}\times[\pi r_{\mathrm{p}}^{2}-k_{\mathrm{ep}}(t)-k_{\mathrm{op}}(t)+k(t)]}
{p_{\mathrm{a}} \pi r_{\mathrm{o}}^{2}+p_{\mathrm{p}} \pi r_{\mathrm{p}}^{2}}\; .
\label{eqt:k}
\end{equation}

Throughout the model, 
the subscript e represents the parameters related to the eclipsing satellite's umbra,  
the subscript o represents the parameters related to the occulting satellite, 
the subscripts a and p represent the parameters related to the active (eclipsing and/or occulting) and passive (eclipsed and/or occulted) satellites respectively. 

\begin{figure}[ht] 
\centering
\includegraphics[width=9 cm]{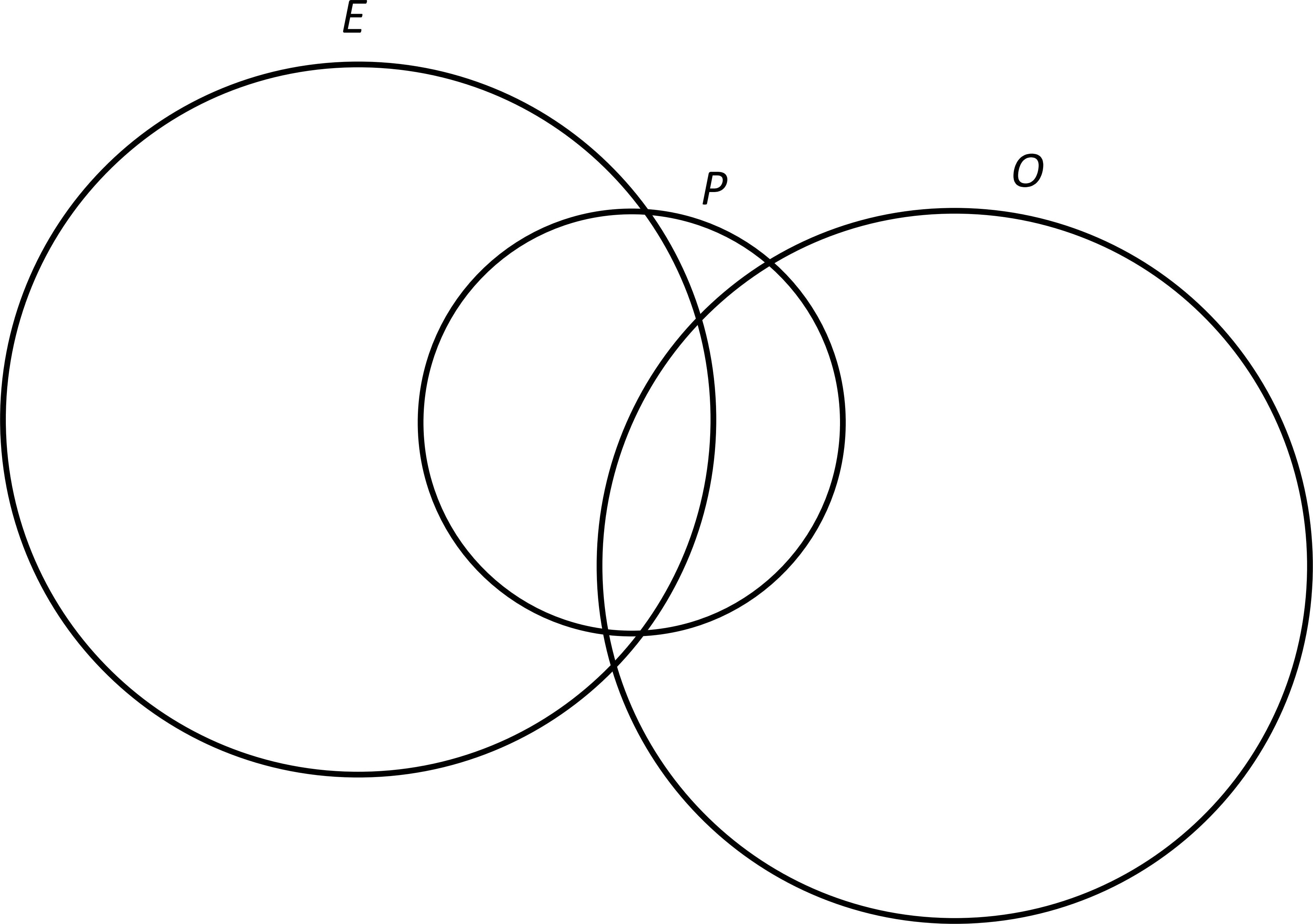}
\caption{Relative positions of the occulting ($O$) and passive ($P$) satellites and the shadow ($E$) of the eclipsing satellite when the passive satellite was eclipsed \textit{AND} occulted simultaneously. \label{fig:uncovered}}
\end{figure}

In Eq.~\ref{eqt:k}, $p$'s and $r$'s are the albedos and the apparent radii respectively. They are constants throughout the event. 
Here, instead of considering a complicated model that involves umbra and penumbra, as well as the gradual decrease of sunlight over the penumbra, we opt to use a simple geometrical model in which the apparent radius of the eclipsing satellite $r_{\mathrm{e}}$ is used as the effective apparent radius of the penumbra. 
Using the formulae in~\citet{emelyanov(2013)}, $r_{\mathrm{e}}$ is $\sim$17\%
larger than the apparent radius of umbra while $r_{\mathrm{e}}$ is $\sim$13\%
smaller than apparent radius of penumbra. In view of the fact that a larger
umbra would overestimate the shadow darkening, on the other hand, a smaller
penumbra would underestimate the shadow darkening, we argue that our
simplification is justifiable.
Since the same satellite acted as the eclipsing and the occulting satellites in our case, $r_{\mathrm{e}}$ equals the apparent radius of the occulting satellite $r_{\mathrm{o}}$.
Specifically, obtained from NASA JPL's \textit{Horizons System},\footnote{\url{https://ssd.jpl.nasa.gov/horizons/app.html}.} 
$r_{\mathrm{e}} = r_{\mathrm{o}} = 0.90535''$ and
$r_{\mathrm{p}} = 0.537''$ is the apparent radius of the passive satellite. 
$k_{\mathrm{ep}}$ is the area covered by the umbra. Similarly, 
$k_{\mathrm{op}}$ is the area covered by the occulting satellite. 
$k$ is the overlapping area among the passive satellite, the active satellite and the umbra. 
For an ordinary ME, $k = 0$ and either $k_{\mathrm{ep}}$ or $k_{\mathrm{op}}$ is to be considered. Note that $[\pi r_{\mathrm{p}}^{2}-k_{\mathrm{ep}}(t)-k_{\mathrm{op}}(t)+k(t)]$ gives the uncovered area of the passive satellite when the passive satellite was eclipsed \textit{AND} occulted simultaneously (see  Fig.~\ref{fig:uncovered}). $S=1$ outside the QSME when all $k$'s are zero.

\begin{figure}[ht] 
\centering
\includegraphics[width=9 cm]{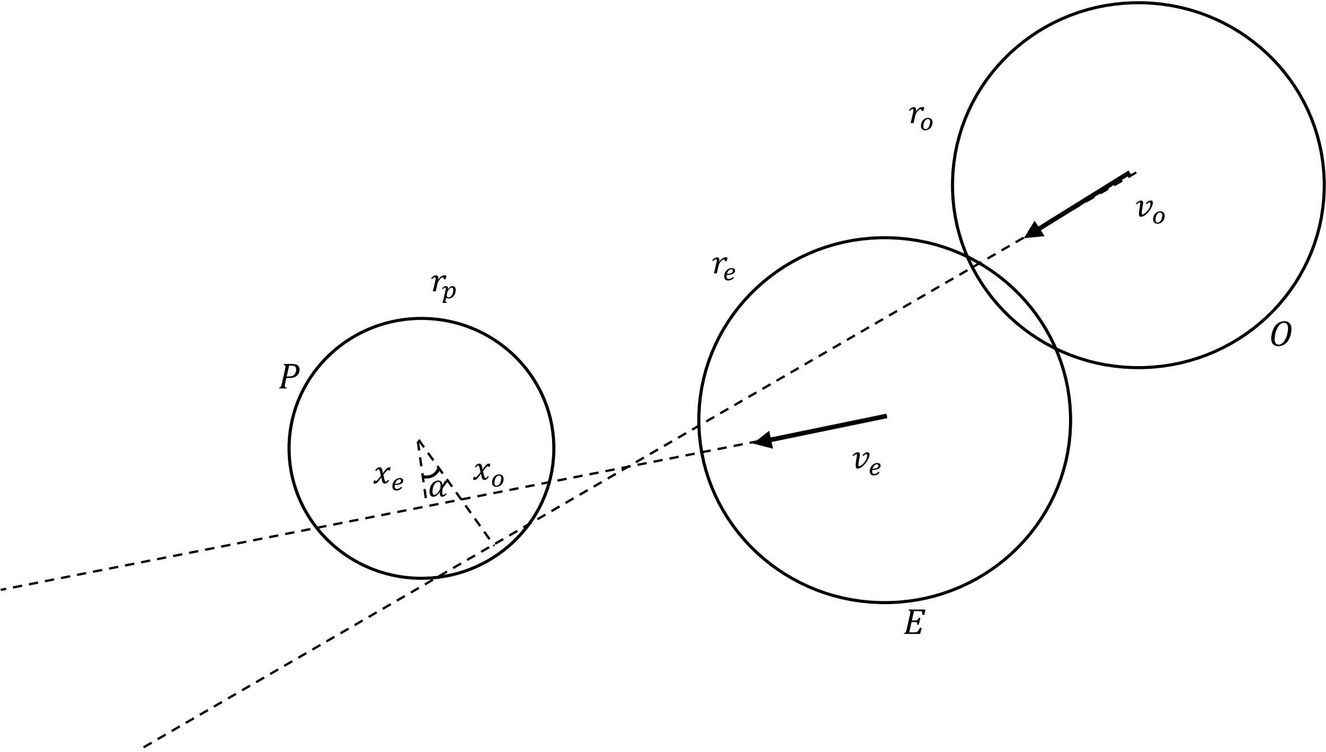}
\caption{Geometry of the QSME model. \label{fig:geometry_before}}
\end{figure}

\begin{figure}[ht] 
\centering
\includegraphics[width=9 cm]{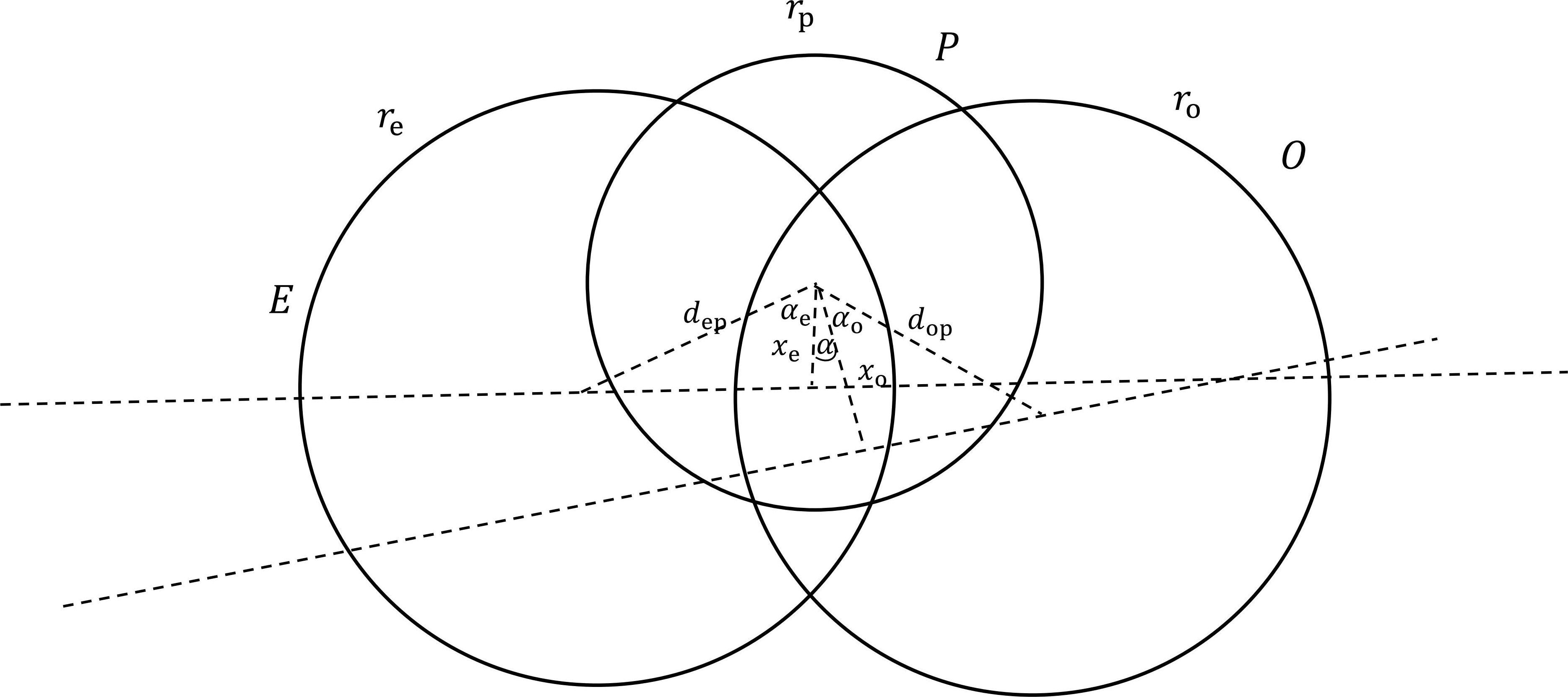}
\caption{Same as Fig.~\ref{fig:geometry_before} when the passive satellite was eclipsed \textit{AND} occulted simultaneously.  
\label{fig:geometry_during}}
\end{figure}

From the astrometric point of view as depicted in Fig.~\ref{fig:geometry_before}, the parameters used to describe the case include: 
\begin{itemize}
\item the impact parameter $x_{\mathrm{e}}$, which is the closest distance between the centres of the circles $P$ (the passive satellite) and $E$ (the umbra of the eclipsing satellite); 
\item the impact parameter $x_{\mathrm{o}}$, which is the closest distance between the centres of the circles $P$ (the passive satellite) and $O$ (the occulting satellite);
\item the satellites' relative speeds $v_{\mathrm{o}}$ and $v_{\mathrm{e}}$; and
\item the angle $\alpha$ between the eclipse and the occultation paths. 
\end{itemize}
Differ from ordinary ME models, $\alpha$ is included because the eclipse and the occultation do not necessarily have to be along the same ``path''. $\alpha_{\mathrm{e}}$ and $\alpha_{\mathrm{o}}$ are also defined as illustrated in Fig.~\ref{fig:geometry_during}.

Here, to model the mutual eclipse and occultation simultaneously, the
satellites' positions are projected onto the plane passing through the centres
of both satellites and the plane is perpendicular to the beam directed to the
satellites from the observer on Earth. For the mutual eclipse, this assumption
differs from ordinary models where the plane is perpendicular to the beam
directed to the satellites from the centre of the Sun~\citep{emelyanov(2020b)}. 
We demonstrate in the phase angle analysis in Appendix E that the astrometric discrepancies originating from the mismatch of projection planes are much smaller than the overall error of each parameter (see 
Table~\ref{tab:results}). 

We also assume that the circles $P$, $E$ and $O$ are perfect, given that the satellites' phase angles are small during the event.

\begin{figure}[ht] 
\centering
\includegraphics[width=9 cm]{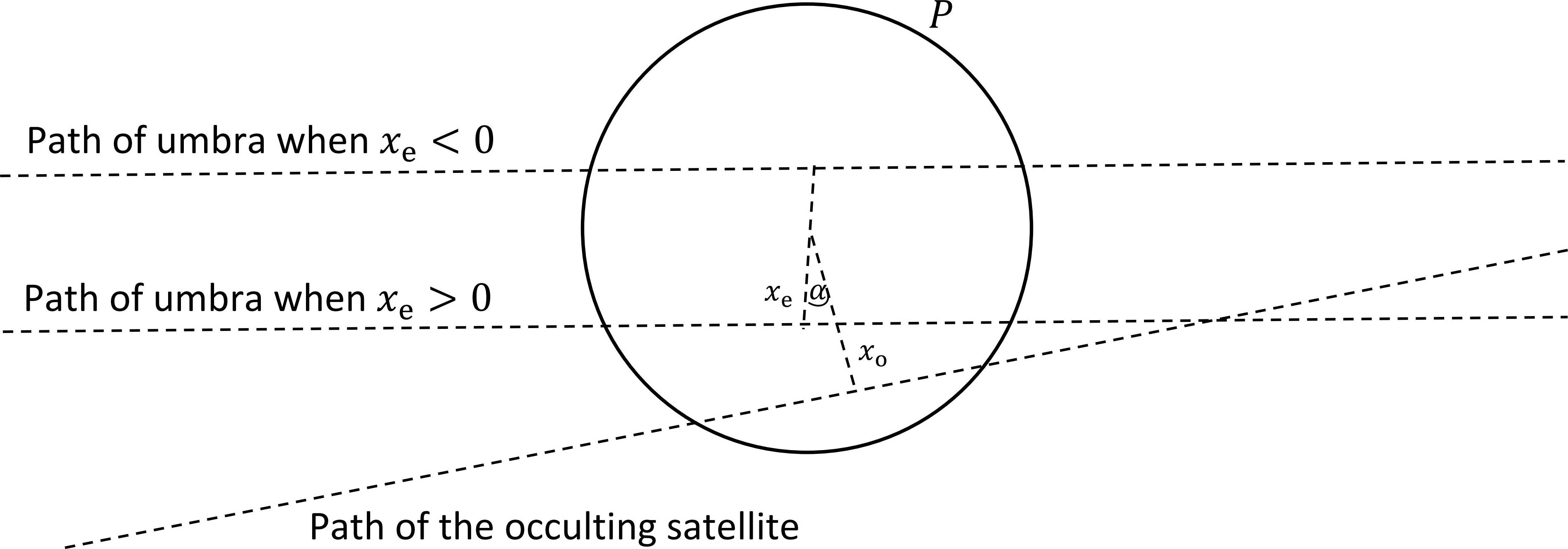}
\caption{Signs of $x_e$ and the path of the umbra relative to the path of the occulting satellite. \label{fig:x_e_sign}}
\end{figure}

According to~\citet{assafin(2009)}, we define $d^{2}=x^{2}+v^{2}\cdot (t-t_{\mathrm{central}})^{2}$, in which $t_{\mathrm{central}}$
is the central time of an individual event, i.e.,~when the eclipse or
occultation is at its maximum. Then at any given time, the distances between
the centres of the circles $P$, $E$ and $O$ can be
obtained from the following three equations:

\begin{equation*}
\begin{aligned}
d_\mathrm{op} &= \sqrt{x_\mathrm{o}^{2}+v_\mathrm{o}^{2}(t-t_\mathrm{o})^{2}} \mbox{\; ,}\\
d_\mathrm{ep} &= \sqrt{x_\mathrm{e}^{2}+v_\mathrm{e}^{2}(t-t_\mathrm{e})^{2}} \mbox{\; and}\\
d_\mathrm{eo} &= \sqrt{d_\mathrm{ep}^{2}+d_\mathrm{op}^{2}-2d_\mathrm{ep}d_\mathrm{op}\cos\beta} \mbox{\; ,}
\end{aligned}
\end{equation*}
where the formulation of $\beta$ depends on the sign of $x_\mathrm{e}$. 
If $x_\mathrm{e}>0$, then $\beta=\alpha +\alpha_\mathrm{e}+\alpha_\mathrm{o}$. 
If $x_\mathrm{e}<0$, i.e.,~the path of the umbra is on the opposite side of the circle,
then $\beta=\pi+\alpha +\alpha_\mathrm{e}+\alpha_\mathrm{o}$ or $\beta=\pi-(\alpha +\alpha_\mathrm{e}+\alpha_\mathrm{o})$ depending on which one is smaller than $\pi$. See Fig.~\ref{fig:x_e_sign}. 
In both cases, 
$\alpha_\mathrm{e}=\tan^{-1}\left[\dfrac{v_\mathrm{e}(t-t_\mathrm{e})}{x_\mathrm{e}}\right]$ and 
$\alpha_\mathrm{o}=\tan^{-1}\left[\dfrac{v_\mathrm{o}(t_\mathrm{o}-t)}{x_\mathrm{o}}\right]$.

The geometric problem becomes expressing $k$'s in terms of $r_{\mathrm{p}}$,
$r_{\mathrm{o}}$, $r_{\mathrm{e}}$, $d_{\mathrm{op}}$, $d_{\mathrm{ep}}$ and $d_{\mathrm{eo}}$, and can
partially be tackled by finding the ``area of common overlap of three
circles''. This has been solved by considering multiple coordinate systems
in~\citet{fewell(2006)} (Appendix A). By geometry, we obtain

\begin{equation*}
\begin{split}
k_\mathrm{op} =  (x''_\mathrm{op}-d_\mathrm{op})\sqrt{r_\mathrm{p}^{2}-(x''_\mathrm{op}-d_\mathrm{op})^{2}}
          +r_\mathrm{p}^{2}\sin^{-1}\left(\dfrac{x''_\mathrm{op}-d_\mathrm{op}}{r_\mathrm{p}}\right)\\
    +r_\mathrm{p}^{2}\dfrac{\pi}{2}+r_\mathrm{o}^{2}\dfrac{\pi}{2}-x''_\mathrm{op}\sqrt{r_\mathrm{o}^{2}-
        (x''_\mathrm{op})^{2}}-r_\mathrm{o}^{2}\sin^{-1} \left(\dfrac{x''_\mathrm{op}}{r_\mathrm{o}}\right)\mbox{\; ,}
\end{split}
\end{equation*}

\begin{equation*}
\begin{split}
k_\mathrm{ep}=({x_\mathrm{ep}'}-d_\mathrm{ep})\sqrt{r_\mathrm{p}^{2}-({x_\mathrm{ep}'}
-d_\mathrm{ep})^{2}}+r_\mathrm{p}^{2}\sin^{-1}\left(\dfrac{{x_\mathrm{ep}'}-d_\mathrm{ep}}{r_\mathrm{p}}\right)\\
+r_\mathrm{p}^{2}\frac{\pi}{2}+r_\mathrm{e}^{2}\frac{\pi}{2}
-{x_\mathrm{ep}'}\sqrt{r_\mathrm{e}^{2}-(x'_\mathrm{ep})^{2}}-r_\mathrm{e}^{2}\sin^{-1}\left(\dfrac{{x_\mathrm{ep}'}}{r_\mathrm{e}}\right)
\end{split}
\end{equation*}

and
\begin{equation*}
\begin{split}
k_\mathrm{eo}=({x_\mathrm{eo}}-d_\mathrm{eo})\sqrt{r_\mathrm{o}^{2}-({x_\mathrm{eo}}
-d_\mathrm{eo})^{2}}+r_\mathrm{o}^{2}\sin^{-1}\left(\dfrac{{x_\mathrm{eo}}-d_\mathrm{eo}}{r_\mathrm{o}}\right)\\
+r_\mathrm{o}^{2}\frac{\pi}{2}+r_\mathrm{e}^{2}\frac{\pi}{2}
-{x_\mathrm{eo}}\sqrt{r_\mathrm{e}^{2}-x_\mathrm{eo}^{2}}-r_\mathrm{e}^{2}\sin^{-1}\left(\dfrac{{x_\mathrm{eo}}}{r_\mathrm{e}}\right)
\mbox{\; .}
\end{split}
\end{equation*}

The details on formulating $k_\mathrm{op}$, $k_\mathrm{ep}$ and $k_\mathrm{eo}$ are provided in Appendix B. The details on formulating $k$, which depends on the configurations of the circles $P$, $E$ and $O$ at a given instant, are provided in Appendix C.

The above formulations assume that both satellites have uniform albedos and
there is no contribution of light reflected from Jupiter (``Jupiter-shine'',
see~\citet{emelyanov(2011)} for the argument). 

We fitted our observations against Eq.~\ref{eqt:measure_model} by the
Orthogonal Distance Regression (ODR) algorithm~\citep{zwolak(2007)} provided in
the data analysis software \textit{OriginLab} 2022 Pro version.\footnote{\url{https://www.originlab.com/2022}.} The fitting
parameters are $x_{\mathrm{e}}$, $x_{\mathrm{o}}$, $v_{\mathrm{e}}$, $v_{\mathrm{o}}$,
$t_{\mathrm{e}}$, $t_{\mathrm{o}}$, $P_{\mathrm{a}}/P_{\mathrm{p}}$, $\alpha$ and $K$. The
next Section presents the fitting results.

\section{Results}
\label{sec:results}

The fitted curve and the O-C residual plot are presented in  Fig.~\ref{fig:light_curve}\textit{(b)} and~\textit{(c)} respectively. As seen from the figures, the fitted curve matches with the observations very well and the absolute values of the residuals are small and no obvious trend is detected. 

We calculated the J3--J2 positional differences in right ascension (RA,
$\alpha$) and declination (DEC, $\delta$), i.e.,~$\Delta \alpha \cos \delta$ and
$\Delta \delta$ respectively, at the $t_{\mathrm{central}}$ from the fitted $x$ and
the position angle obtained from the~\citet{lainey(2009)}'s theory with the
INPOP19a planetary ephemerides~\citep{fienga(2020)} on
\textit{MULTI-SAT}.\footnote{\url{http://nsdb.imcce.fr/multisat/nssreq5he.htm}.}
The corresponding O-C's are presented in 
Table~\ref{tab:position_diff}.
The positional differences show a good agreement with the theory. Particularly,
the sizes of our O-C's are very similar to those of previous ME studies
reported for the last four observation campaigns (PHEMU03, PHEMU09, PHEMU15 and
PHEMU21, see~\citealp{arlot(2019),emelyanov(2022)}). Notably, our occultation
results have smaller discrepancies.

\begin{table}[ht]
\centering
\caption{O–C's of satellite-satellite positional differences in milliarcseconds (mas).
\label{tab:position_diff}}
\centering
\begin{tabular}{rrrrrr}
\hline
\textbf{Date} & \textbf{Event} & \textbf{Model} & \textbf{Theory, ephemerides} & \textbf{O-C (RA)} & \textbf{O-C (DEC)}\\
\hline
22 August 2021  & 3E2 & present study & \citet{lainey(2009),fienga(2020)} & -25.9 & 30.3 \\ 
22 August 2021  & 3O2 & present study & \citet{lainey(2009),fienga(2020)} & -6.2 & 17.9 \\ 
29 January 1991 & 2O1 & present study & \citet{lainey(2009),fienga(2020)} & 36.0 & 55.3 \\ 
29 January 1991 & 2E1 & present study & \citet{lainey(2009),fienga(2020)} & 0.5 & 0.5 \\ 
29 January 1991 & 2O1 & \citet{vasundhara(1994)} & \citet{lieske(1977),lieske(1987)} & -61.9 & -204.1 \\ 
29 January 1991 & 2E1 & \citet{vasundhara(1994)} & \citet{lieske(1977),lieske(1987)} & 26.1 & 87 \\ 
\hline
\end{tabular}
\end{table}

We derived the beginning time, ending time and flux drop ratio from the fitted model. 
Table~\ref{tab:results} shows the fitted and derived parameters and the O-C residuals for the parameters predicted by \textit{MULTI-SAT} in 
Table~\ref{tab:prediction}. 
In addition, the O-C's of satellites' relative speeds were compared with those computed from INPOP19a.
Since the satellites' phase angles during the QSME were very small
($0.680^{\circ}$ -- $0.695^{\circ}$), the O-C of satellites' albedos ratio was
compared with those at $0^{\circ}$ phase angle listed on the
\textit{MULTI-SAT}'s
website\footnote{\url{http://nsdb.imcce.fr/multisat/parcohe.htm}.}
($P_{\mathrm{J3}}/P_{\mathrm{J2}} = 0.43/0.64 = 0.672$). The comparison takes the ``opposition surge'', i.e.~the surge in
brightness when the object has a very small phase
angle~\citep{fujii(2014),buratti(1995)}, into account.
As noted in Sect.~\ref{sec:model} and detailed in Appendix E, we computed the O-C of 3E2's parameters even though the projection planes of our model and those used in \textit{MULTI-SAT}'s predictions were different.
The fitting yields a negative $x_{\mathrm{e}}$. We report the absolute value of $x_{\mathrm{e}}$ in 
Table~\ref{tab:results} and compare the value with the prediction.

\begin{table*}[ht]
\centering
\caption{Results of the model fitted to the observations of the QSME event on 22 August 2021. O-C residuals refer to the predictions by \textit{MULTI-SAT} (Table~\ref{tab:prediction}). See the text for remarks on the eclipse's O-C's.
\label{tab:results}}
\centering
\begin{tabular}{lrrr}
\hline
\textbf{Parameter} & \textbf{Description (unit)}       & \textbf{Fitted value $\pm$ standard error}  & \textbf{O-C}\\
\hline
\multicolumn{4}{c}{\textbf{3E2}} \\
$x_\mathrm{e}$ & impact parameter (arcsec)                    &  0.103 $\pm$ 0.119              &  0.078 \\
$v_\mathrm{e}$ & relative speed at $t_\mathrm{e,central}$ (arcsec hr$^{-1}$)      &  2.855 $\pm$ 0.058              &  0.175 \\
$t_\mathrm{e,begin}$ & beginning time (hour after 13:00)      &  1.018 $\pm$ 0.016              &  0.034 \\
$t_\mathrm{e,central}$       & central time (hour after 13:00)        &  1.525 $\pm$ 0.011              &  0.012 \\
$t_\mathrm{e,end}$   & ending time (hour after 13:00)         &  2.031 $\pm$ 0.016              & -0.013 \\
$\Delta I_\mathrm{e}$ & flux drop ratio                       &  0.639 $\pm$ 0.001              & -0.014 \\
\hline
\multicolumn{4}{c}{\textbf{3O2}} \\
$x_\mathrm{o}$ & impact parameter (arcsec)                    &  0.386 $\pm$ 0.004              & -0.019 \\
$v_\mathrm{o}$ & relative speed at $t_\mathrm{o,central}$ (arcsec hr$^{-1}$)      &  3.153 $\pm$ 0.026              &  0.012 \\
$t_\mathrm{o,begin}$ & beginning time (hour after 13:00)      &  1.529 $\pm$ 0.005              & -0.036 \\
$t_\mathrm{o,central}$       & central time (hour after 13:00)        &  2.003 $\pm$ 0.002              & -0.004 \\
$t_\mathrm{o,end}$   & ending time (hour after 13:00)         &  2.477 $\pm$ 0.005              &  0.029 \\
$\Delta I_\mathrm{o}$ & flux drop ratio                       &  0.641 $\pm$ 0.001              & -0.017 \\
\hline
\multicolumn{4}{c}{\textbf{QSME}} \\
$P_\mathrm{a}/P_\mathrm{p}$ & ratio of satellites’ albedos            &   0.624  $\pm$ 0.001         & -0.048\\
$\alpha$ & eclipse-occultation path separation angle (radian)   &   -0.210 $\pm$ 0.055         & $\cdots$ \\
$K$ & proportionality constant                                  &   2.161  $\pm$ 0.001         & $\cdots$ \\ 
\hline
\end{tabular}
\end{table*}

As a visual validation, we created an animated simulation based on the fitted parameters in \url{https://www.geogebra.org/calculator/rx6efjeu}. 

Previous studies analysed multiple light curves obtained from different ordinary MEs and published the O-C's for each event. 
We compared our O-C values on the flux drop ratio ($-2.1\%$ and $-2.6\%$ for the mutual eclipse and occultation respectively) with the ranges of O-C of previous studies. 
The measured O-C's of flux drops reported in~\citet{peng(2007)} ranged from
$-4\%$ to 9\% (excluding one observation with an O-C valued
$-45\%$). 
Similarly, the O-C's reported in~\citet{vienne(2003)} ranged from $-26\%$
to 10\%. 
\citet{zhang(2019)} even reported a O-C of mutual eclipse magnitude drop as
large as $-80\%$ and a O-C of mutual occultation magnitude drop as large
as 2100\%.  

We also compared our O-C values on the central time (43.2 and $-14.4$~s for the mutual eclipse and occultation respectively) with the ranges of O-C of previous studies. 
Based on the predictions from~\citet{lainey(2009)}'s theory,~\citet{dias(2013)}
reported that the O-C of mutual eclipses ranged from $-41.6$ to 22~s
while that of mutual occultations ranged from $-10.9$ to 19.2~s.   
Based on the computations of E3
ephemerides~\citep{lieske(1977),lieske(1987)},~\citet{vasundhara(1994)}
reported that the O-C of mutual eclipses ranged from $-6.9$ to
36.1~s while that of mutual occultations ranged from 18.0 to 37.5~s
(excluding one observation with poor data quality).

In sum, we conclude that the absolute values of O-C in the present study are comparable to or even smaller than those of the previous studies in most of the cases. In other words, our simple geometric model is reasonably accurate with the high-quality data obtained in our observations.

\subsection{Separate the light curve into an eclipse and an occultation counterparts}
\label{sec:individual_fit}
To validate the hypothesis that the observed QSME is a superposition of the eclipse and the occultation counterparts, we attempted to separate the QSME light curve into individual counterparts, based on the assumption that the eclipse and occultation light curves are symmetric before and after the corresponding central time ($t_{\mathrm{e,central}}$ and $t_{\mathrm{o,central}}$).

\begin{figure*}[ht] 
\centering
\includegraphics[width=16 cm]{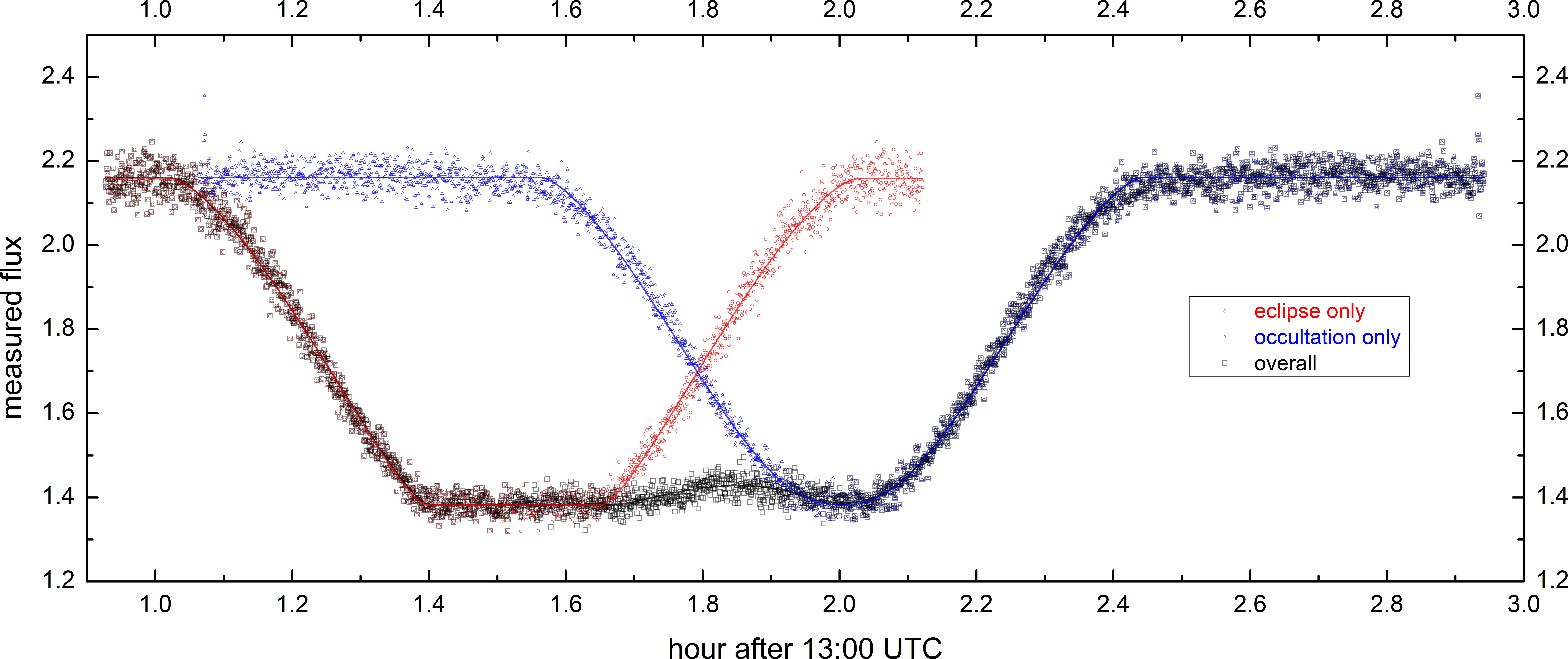}
\caption{Separation of the eclipse and the occultation counterparts of the QSME event on 22 August 2021. The fitted light curves are overlaid on the data points. See the text for details. \label{fig:fit_separate}}
\end{figure*}

First, we reproduced the second half of the 3E2 light curve by mirroring the observations before the fitted $t_{\mathrm{e,central}}$ (see 
Table~\ref{tab:results}).
Second, we fitted the complete 3E2 light curve against the model presented in  Fig.~\ref{sec:model}. The model was simplified to include the eclipse part only, i.e.,~setting $k_{\mathrm{op}} = k = 0$ in Eq.~\ref{eqt:k}. 
Similarly for the occultation, we reproduced the complete 3O2 light curves by mirroring the observations after the fitted $t_{\mathrm{o,central}}$ (see 
Table~\ref{tab:results}) and conducted the fitting against the simplified occultation-only model, i.e.,~setting $k_{\mathrm{ep}} = k = 0$ in Eq.~\ref{eqt:k}. 
Fig.~\ref{fig:fit_separate} shows the half-observed-half-reproduced light curves and their fitted curves. 
Table~\ref{tab:individual_fit_results} tabulates the fitting results and O-C comparisons.  

\begin{table*}[ht]
\centering
\caption{Separation and fitting of the eclipse and the occultation counterparts of the QSME event on 22 August 2021. O-C residuals refer to the predictions by \textit{MULTI-SAT} (Table~\ref{tab:prediction}). See Table~\ref{tab:results} for the parameters' units.
\label{tab:individual_fit_results}}
\centering
\begin{tabular}{crrr}
\hline
\textbf{         }          & \textbf{3E2 only}              & \textbf{3O2 only}              & \textbf{}\\
\textbf{Parameter}          & \textbf{Fitted value $\pm$ standard error} & \textbf{Fitted value $\pm$ standard error} & \textbf{O-C}\\
\hline
$x_\mathrm{e}$                       & 0.005 $\pm$ 0.436            & $\cdots$                         &  -0.020 \\
$v_\mathrm{e}$                       & 2.880 $\pm$ 0.010            & $\cdots$                         &  0.200 \\
$\Delta I_\mathrm{e}$                &  0.640 $\pm$ 0.001            & $\cdots$                         & -0.012 \\
$x_\mathrm{o}$                       & $\cdots$                     &  0.401 $\pm$ 0.022                &  -0.005 \\
$v_\mathrm{o}$                       & $\cdots$                     & 3.157 $\pm$ 0.019                &  0.016 \\
$\Delta I_\mathrm{o}$                & $\cdots$                     & 0.639 $\pm$ 0.008                & -0.018 \\
$P_\mathrm{a}/P_\mathrm{p}$        & 0.626 $\pm$  0.002              & 0.610 $\pm$ 0.016                & -0.046 \& -0.062 \\
\hline
\end{tabular}
\end{table*}

As seen from Fig.~\ref{fig:fit_separate} and the results in 
Table~\ref{tab:individual_fit_results}, the fittings are good and the absolute values of O-C's are very small in general. The separation of the QSME light curve has been successful. 

Relatively large O-C discrepancies are only found on $v_{\mathrm{e}}$ from both the combined (6.5\%, see 
Table~\ref{tab:results}) and the separated (8.1\%, see 
Table~\ref{tab:individual_fit_results}) fittings. We attribute the large discrepancies to our model's incapability of including the satellites' acceleration. Future works are required to take the acceleration, although small, into consideration. 

\subsection{Fitting the 2O1 + 2E1 observation on 29 January 1991}
\label{sec:fit1991}

\begin{figure*}[ht] 
\centering
\includegraphics[width=16 cm]{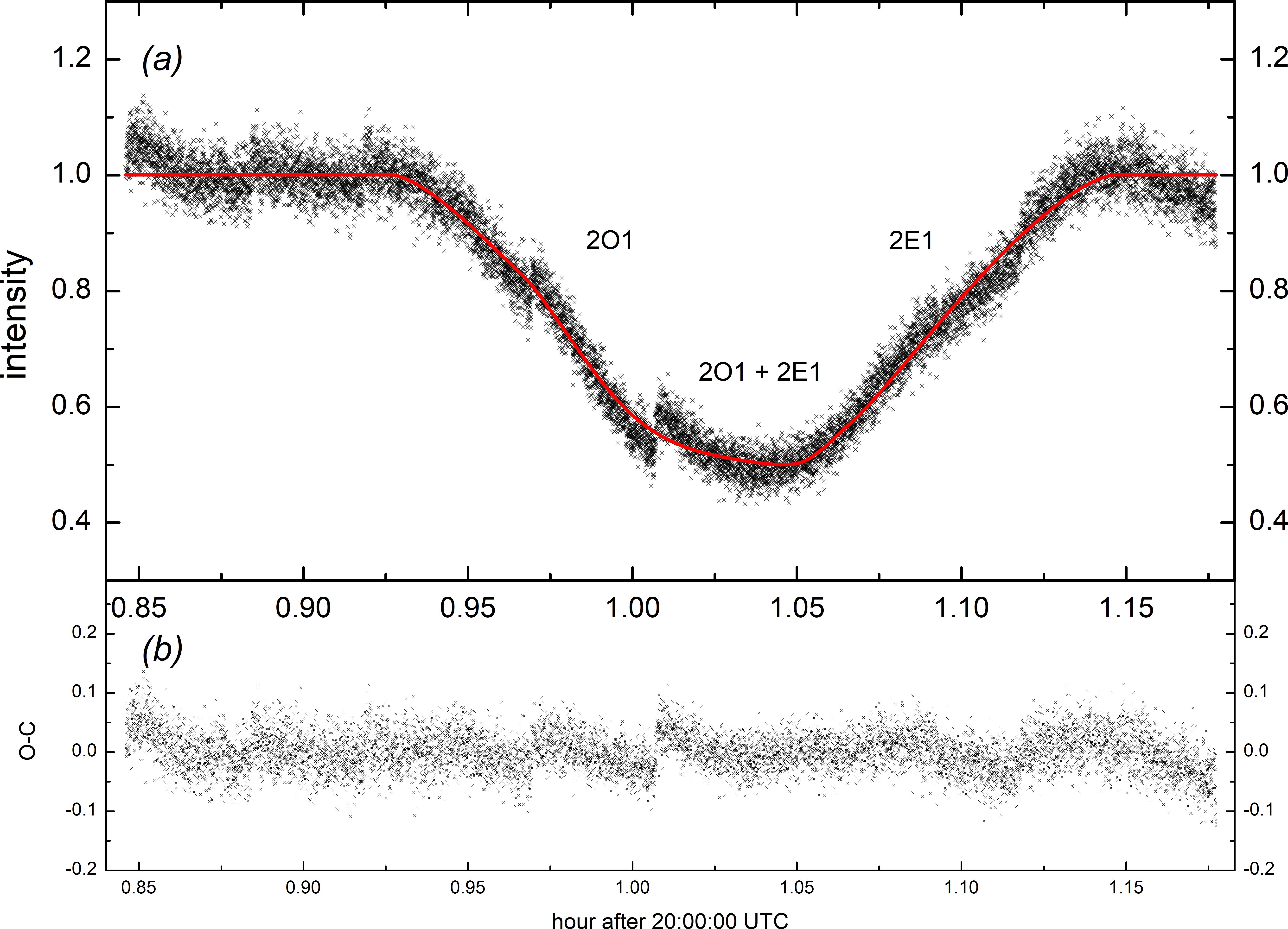}
\caption{The 2O1 + 2E1 event on 29 January 1991.
\textit{(a)}: observed data points (by others) and the fitted light curve in red; 
\textit{(b)}: O-C. 
\label{fig:light_curve_1991}}
\end{figure*}

As mentioned in Sect.~\ref{sec:introudction} and simulated in  Fig.~\ref{fig:QSME_examples}\textit{(c)}, a QSME of 2O1 $+$  2E1 happened on 29 January 1991. The database of the Institut de M\'ecanique C\'eleste et de Calcul des Eph\'em\'erides (IMCCE)'s Natural Satellites Service archives several sets of photometric measurements of the event. We fitted the one with the best data quality taken from the Observatory at Calern in France (\textit{dgrav2oe1.291} channel V dataset, converted to linear intensity) with the present model.
The fitted asymmetric light curve and the O-C residual plot are presented in  Fig.~\ref{fig:light_curve_1991}\textit{(a)} and~\textit{(b)} respectively. 
Table~\ref{tab:results_1991} shows the fitted parameters and their errors. The fitting is very good and yields a negative $x_{\mathrm{e}}$. We report the absolute value of $x_{\mathrm{e}}$ in 
Table~\ref{tab:results_1991} and compare the value with the prediction.

\begin{table*}[ht]
\centering
\caption{Results of our model fitted to the observations of the QSME event by others on 29 January 1991. \label{tab:results_1991}}
\centering
\begin{tabular}{lrrr}
\hline
\textbf{Parameter} & \textbf{Description (unit)}       & \textbf{Fitted value $\pm$ standard error}  & \textbf{O-C}\\
\hline
\multicolumn{4}{c}{\textbf{2O1}} \\
$x_\mathrm{o}$ & impact parameter (arcsec)                    &  0.395 $\pm$ 0.021              & 0.068 \\
$v_\mathrm{o}$ & relative speed at $t_\mathrm{o,central}$ (arcsec hr$^{-1}$)      &  14.107 $\pm$ 0.096             &  1.621 \\
$t_\mathrm{o,central}$       & central time (hour after 20:00)        &  0.999 $\pm$ 0.001              & 0.001 \\
\hline
\multicolumn{4}{c}{\textbf{2E1}} \\
$x_\mathrm{e}$ & impact parameter (arcsec)                    &  0.080 $\pm$ 0.000             &  0.029 \\
$v_\mathrm{e}$ & relative speed at $t_\mathrm{e,central}$ (arcsec hr$^{-1}$)      &  12.150 $\pm$ 0.095             &  2.038\\
$t_\mathrm{e,central}$       & central time (hour after 20:00)        &  1.057 $\pm$ 0.001              & 0.011 \\
\hline
\multicolumn{4}{c}{\textbf{QSME}} \\
$P_\mathrm{a}/P_\mathrm{p}$ & ratio of satellites’ albedos            &  0.871 $\pm$ 0.015        & -0.145\\
$\alpha$ & eclipse-occultation path separation angle (radian)   &   -0.505 $\pm$ 0.071        & $\cdots$ \\
$K$ & proportionality constant                                  &  1.000 $\pm$ 0.001       & $\cdots$ \\ 
\hline
\end{tabular}
\end{table*}

Similarly to our 2021 observations, we present the comparisons between theoretical J2--J1 positional differences and our calculations in 
Table~\ref{tab:position_diff}. O-C's published by~\citet{vasundhara(1994)} on the
same event (but different observations) are appended for reference. Our method
is not only applicable to the 1991 observations but also produces results that
are closer to the theoretical values.

\section{Conclusions \& Discussions}
\label{sec:conclusions}

Previous works focused primarily on separated MEs. Under specific rare orbital configurations when the same satellite is eclipsed and occulted quasi-simultaneously, a QSME occurs. Few researchers observed QSMEs and even fewer attempted to analyse their light curves. The present study fills the gap by observing a QSME on 22 August 2021, analysing the light curve and developing a model to obtain astrometric parameters. 

Our study demonstrated that the QSME can be observed as easily as ordinary MEs. The QSME created an asymmetric and long ($\sim$1.5~h) light curve, which represents the superposition of an eclipse and an occultation counterparts. The light curve exhibited fine details such as the small amount of light escaped from the uncovered area of the near-total occulted satellite. 

We also demonstrated that formulating the QSME by a simple geometric model is feasible.
Despite the fact that high-order factors such as the effects of penumbra, satellites' surface albedo variations and Jupiter's halo had not been considered, the model successfully explained the entire QSME, with the absolute values of O-C's comparable to those of previous ME studies which adopted complicated models. Our simple model performs well as it describes two inter-correlated MEs (an eclipse \textit{AND} an occultation) with improved sensitivity on the fitted parameters with respect to the observations on the QSME than those on individual MEs. Particularly in our case, the shape of the light curve during the period when the eclipse and the occultation occurred simultaneously places an additional constraint on the eclipse and the occultation parameters.
In other words, obtaining more accurate results from QSME observations is possible even with a model that is not sophisticated. As demonstrated by fitting the 1991 QSME observations in Fig.~\ref{sec:fit1991}, our model is generic and can be easily modified to apply to other cases. Future works include further testing our model with other observers' data (past or future). 

Rough searches from the predictions of \textit{MULTI-SAT} database\footnote{\url{http://nsdb.imcce.fr/multisat/nsszph5he.htm}.} and \textit{Occult} software reveal that the Galilean satellites' orbital configurations do not favour any QSME during the next observing season in 2026--27.
From \textit{Occult}, the next QSME (2E3 $+$ 2O3) occurring  04:42--04:56, 3
February 2033 seems to be the only QSME in the 2032--2033 observing season.
Unfortunately, Jupiter will be very close ($\sim$39') to the Sun at that
time because the Sun--Jupiter conjunction happens on 2 February, although
daylight infrared ME observations have been shown possible~\citep{arlot(1990)}.

\begin{figure}[ht] 
\centering
\includegraphics[width=8 cm]{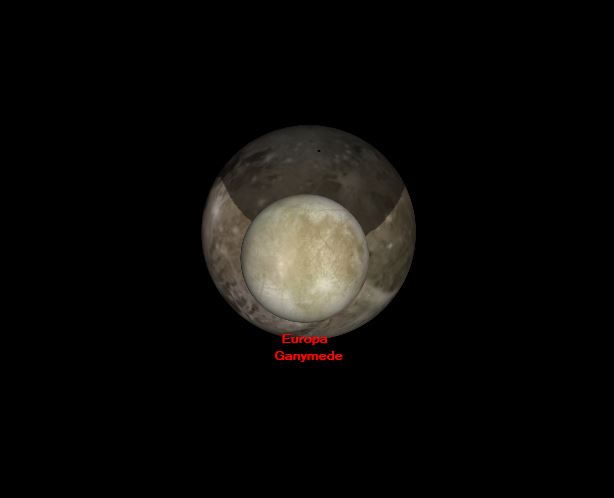}
\caption{Next visible QSME (4E3 + 2O3) will occur 19:09--19:35, 6 January 2045. The shadow belongs to J4 instead of the occulting satellite J2. This event kick-starts a few more QSMEs in the same year. Simulations created with WinJUPOS (\url{http://www.grischa-hahn.homepage.t-online.de/})\label{fig:2045_QSME}}
\end{figure}

The next truly visible QSME (4E3 $+$ 2O3 as simulated in Fig.~\ref{fig:2045_QSME}) will occur on 6 January 2045 . Indeed, 2045 is an unusual year in which at least four more QSMEs will happen in a multitude of different ways.\footnote{2O3 $+$  2E3, 00:07--00:22, 6 February 2045; 2O1 $+$  2E1, 02:21--02:28, 7 February 2045; 3E2 $+$  3O2, 15:40--15:49, 8 February 2045; 2E1 $+$  2O1, 15:32--15:44, 10 February 2045. Occurrences are predicted by \textit{Occult} software.} QSMEs in 2045 collectively involve all Galilean satellites. 

Given the rarity, our observation in 2021 remains unique in the short future.

The key to improving our knowledge of Solar System dynamics relies on accurate
astrometric solutions reduced from many different ME observations covering
extended periods~\citep{emelyanov(2020b),arlot(2019),arlot(2009b)}.  
Given that QSME observations serve the same purpose as demonstrated in the present work, we encourage professional and amateur observers to observe QSMEs in 2045 and beyond. 
In the meantime, we strongly encourage observers to release previous and future photometric data on QSMEs as much as possible to enrich the existing small database for examination and analysis.

\section*{Data Availability}
The photometric measurements and astrometric results in the present study have be uploaded to the database of IMCCE's Natural Satellites Service. 

\section*{Declaration of Competing Interest}
None.

\section*{Acknowledgements} 
The authors thank Professor Ming-chung Chu for helpful discussions on the data analysis. G. Luk and G. Chung received support from the Department of Physics of The Chinese University of Hong Kong. E. Yuen received support from the Department of Physics of The University of Hong Kong. We would also like to thank reviewers for taking the time and effort necessary to review the manuscript.

\bibliographystyle{cas-model2-names}
\bibliography{QSME_v7}

\section*{Appendix A: Coordinate Systems\label{app:coordinate}}

\normalsize

\begin{figure}[ht] 
\centering
\includegraphics[width=9 cm]{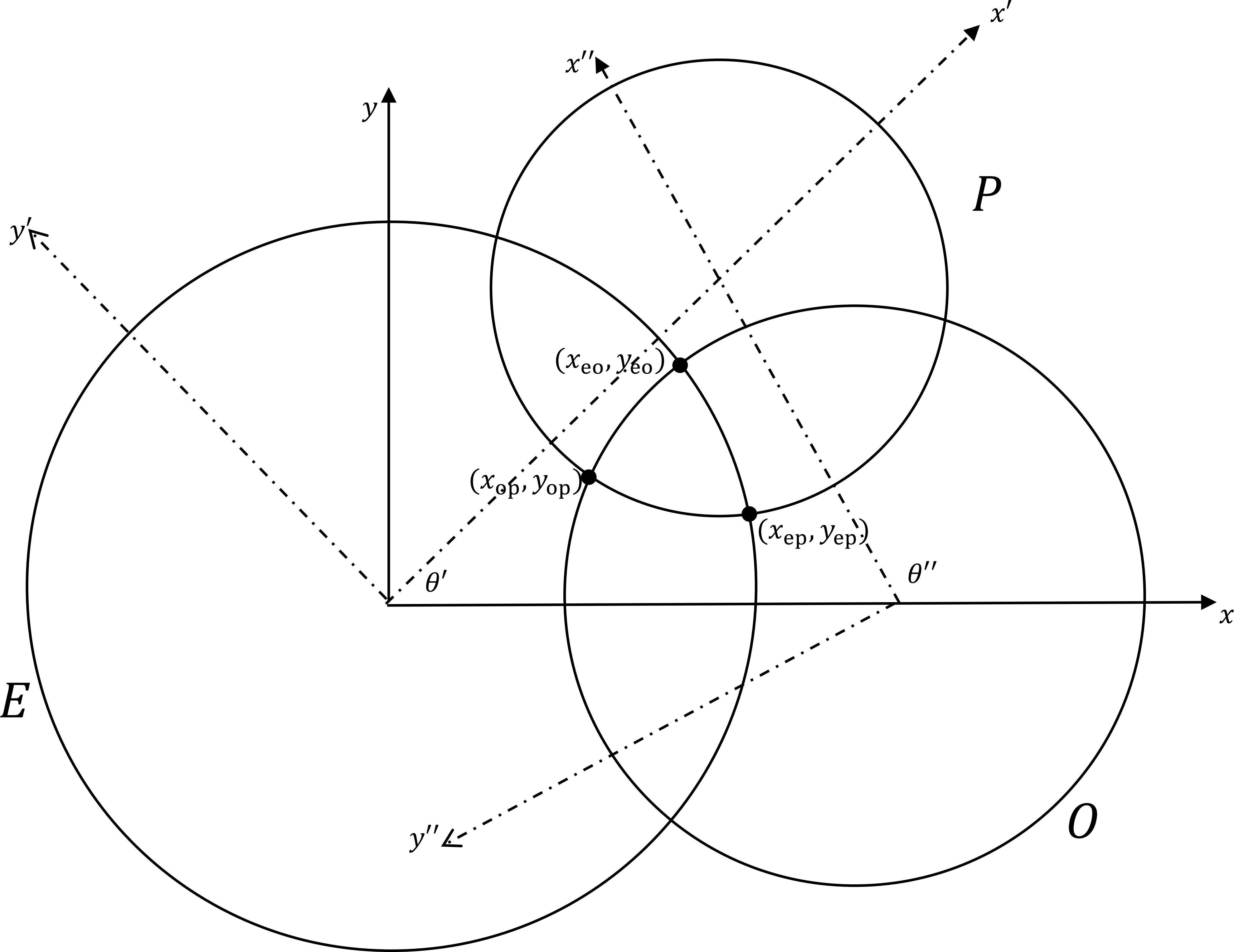}
\caption{Coordinate systems for finding the area of common overlap of three circles. \label{fig:coordinate_system}}
\end{figure}

The coordinate systems follow~\citet{fewell(2006)}. As illustrated in
Fig.~\ref{fig:coordinate_system}, the origin of coordinates is placed at the centre
of the circle $E$ and the $x$-axis passes through the centre
of the circle $O$. The direction of the $y$-axis is chosen so
that the $y$-coordinate of the centre of the circle $P$ is
positive.

The $(x, y)$ system is appropriate for analysing the intersections of the circles $E$ and $O$. To describe the third circle, for convenience, two other coordinate systems are used. They are labelled as $(x', y')$ and $(x', y')$ and obtained by rotating and translating the $(x, y)$ coordinates:

\begin{itemize}
\item The origin of the $(x',y')$ system locates at the centre of the circle $E$ and the $x'$ axis passes through the centre of the circle $P$.
\item The origin of the $(x',y')$ system locates at the centre of the circle $O$ and the $x'$ axis passes through the centre of the circle $P$.

\end{itemize}

Fig.~\ref{fig:coordinate_system} also shows the angles $\theta'$ and $\theta''$ between the $x$-axis and the respective abscissas of the two additional systems.

Before finding $k_\mathrm{op}$, $k_\mathrm{ep}$ and $k$, let’s find the three intersections among circles, namely $(x_\mathrm{eo},y_\mathrm{eo})$, $(x_\mathrm{ep},y_\mathrm{ep})$ and $(x_\mathrm{op},y_\mathrm{op})$. The first set is:

\begin{equation*}
\begin{aligned}
x_\mathrm{eo} &= \dfrac{r_\mathrm{e}^{2}-r_\mathrm{o}^{2}+d_\mathrm{eo}^{2}}{2d_\mathrm{eo}} \mbox{\; and} \\
y_\mathrm{eo} &= \dfrac{1}{2d_\mathrm{eo}}\sqrt{2d_\mathrm{eo}^{2}(r_\mathrm{e}^{2}+r_\mathrm{o}^{2})-(r_\mathrm{e}^{2}-
          r_\mathrm{o}^{2})^{2}-d_\mathrm{eo}^{4}}\mbox{\; .}
\end{aligned}
\end{equation*}

To find $(x_\mathrm{ep}, y_\mathrm{ep})$, we need to find $\theta'$ using the cosine formula and $(x'_\mathrm{ep}, y'_\mathrm{ep})$, which is the coordinate of the point in $(x',y')$. Similarly, to find $(x_\mathrm{op}, y_\mathrm{op})$, we need to find $\theta''$ using the cosine formula and $(x''_\mathrm{op}, y''_\mathrm{op})$, which is the coordinate of the point in $(x'',y'')$. Specifically, related equations are: 

\begin{equation*}
\begin{aligned}
\cos\theta' &= \dfrac{d_\mathrm{eo}^{2}+d_\mathrm{ep}^{2}-d_\mathrm{op}^{2}}{2d_\mathrm{eo}d_\mathrm{ep}} \mbox{\; ,}\\
\sin\theta' &= \sqrt{1-\cos^{2}\theta'} \mbox{\; ,}\\
x'_\mathrm{ep} &= \dfrac{r_\mathrm{e}^{2}-r_\mathrm{p}^{2}+d_\mathrm{ep}^{2}}{2d_\mathrm{ep}} \mbox{\; ,}\\
y'_\mathrm{ep} &= -\dfrac{1}{2d_\mathrm{ep}}\sqrt{2d_\mathrm{ep}^{2}(r_\mathrm{e}^{2}+r_\mathrm{p}^{2})-(r_\mathrm{e}^{2}-
r_\mathrm{p}^{2})^{2}-d_\mathrm{ep}^{4}} \mbox{\; ,}\\
\cos\theta'' &= -\dfrac{d_\mathrm{eo}^{2}+d_\mathrm{op}^{2}-d_\mathrm{ep}^{2}}{2d_\mathrm{eo}d_\mathrm{op}} \mbox{\; ,}\\
\sin\theta'' &= \sqrt{1-\cos^{2}\theta''} \mbox{\; ,}\\
x_\mathrm{op}'' &= \dfrac{r_\mathrm{o}^{2}-r_\mathrm{p}^{2}+d_\mathrm{op}^{2}}{2d_\mathrm{op}} \mbox{\; and}\\
y_\mathrm{op}'' &= \dfrac{1}{2d_\mathrm{op}}\sqrt{2d_\mathrm{op}^{2}(r_\mathrm{o}^{2}+r_\mathrm{p}^{2})-(r_\mathrm{o}^{2}-r_\mathrm{p}^{2})^{2}-d_\mathrm{op}^{4}}\mbox{\; .}
\end{aligned}
\end{equation*}

Now we can transform the point $(x'_\mathrm{ep}, y'_\mathrm{ep})$ and $(x''_\mathrm{ep}, y''_\mathrm{ep})$ back to the system $(x, y)$, i.e.,

\begin{equation*}
\begin{aligned}
x_\mathrm{ep} &= x'_\mathrm{ep}\cos\theta'-y_\mathrm{ep}'\sin\theta' \mbox{\; ,}\\
y_\mathrm{ep} &= x'_\mathrm{ep}\sin\theta'+y_\mathrm{ep}'\cos\theta' \mbox{\; ,}\\
x_\mathrm{op} &= x''_\mathrm{op}\cos\theta''-y''_\mathrm{op}\sin\theta''+d_\mathrm{eo} \mbox{\; and}\\
y_\mathrm{op} &= x''_\mathrm{op}\sin\theta''+y_\mathrm{op}''\cos\theta''\mbox{\; .}
\end{aligned}
\end{equation*}

\section*{Appendix B: Formulating $k_{\rm op}(t)$, $k_{\rm ep}(t)$ and $k_{\rm eo}(t)$ \label{app:k_op_ep_eo}}

\begin{figure}[ht] 
\centering
\includegraphics[width=9 cm]{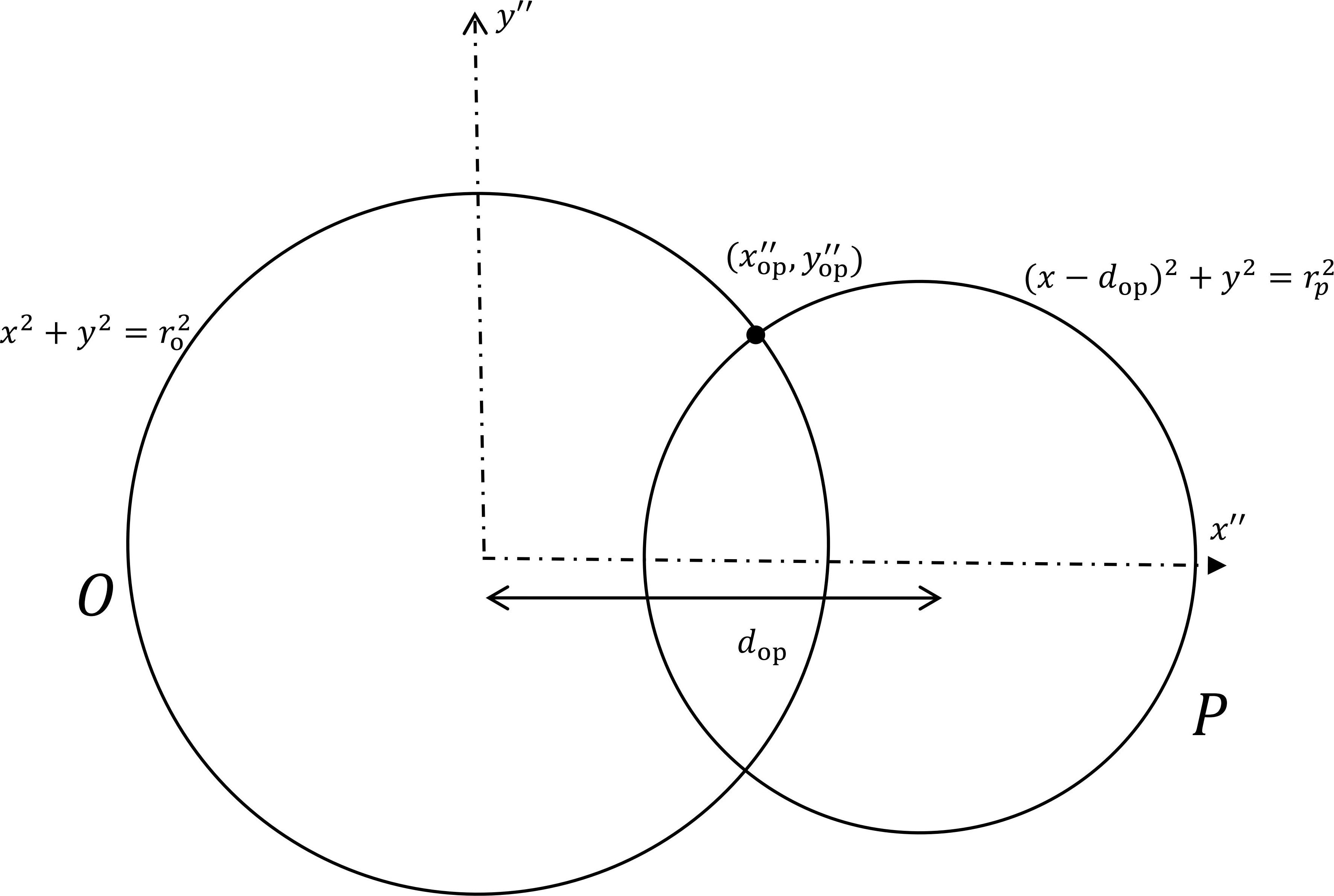}
\caption{$(x'',y'')$ coordinate system when finding $k_\mathrm{op}(t)$. \label{fig:dop_coordinate}}
\end{figure}

$k_\mathrm{op}(t)$ can be found by integration in the coordinate system $(x'', y'')$ (Fig.~\ref{fig:dop_coordinate}). Note that for $d_\mathrm{op} >r_\mathrm{p}+r_\mathrm{o}$, $k_\mathrm{op}(t)=0$ since satellites do not overlap with each other. $k_\mathrm{op}(t)=1$ for the case that the passive satellite is completely occulted, i.e., $d_\mathrm{op}<r_\mathrm{o}-r_\mathrm{p}$. 

\begin{equation*}
\begin{aligned}
k_\mathrm{op}(t) &= 2\int^{x''_\mathrm{op}}_{d_\mathrm{op}-r_\mathrm{p}}\sqrt{r_\mathrm{p}^{2}-(x-d_\mathrm{op})^{2}}
           dx +2\int^{r_\mathrm{o}}_{x''_\mathrm{op}}\sqrt{r_\mathrm{o}^{2}-x^{2}}dx\\
       &= 2\int^{x''_\mathrm{op}-d_\mathrm{op}}_{-r_\mathrm{p}}\sqrt{r_\mathrm{p}^{2}-u^{2}}
           du +2\int^{r_\mathrm{o}}_{x''_\mathrm{op}}\sqrt{r_\mathrm{o}^{2}-x^{2}}dx \\
       &= \left[ u\sqrt{r_\mathrm{p}^{2}-u^{2}}+r_\mathrm{p}^{2}
          \sin^{-1}\left(\dfrac{u}{r_\mathrm{p}}\right)\right]^{x''_\mathrm{op}-d_\mathrm{op}}_{-r_\mathrm{p}}\\
       &\;\;\;\;+\left[ x\sqrt{r_\mathrm{o}^{2}-x^{2}}+r_\mathrm{o}^{2}\sin^{-1}\left(\dfrac{x}{r_\mathrm{o}}\right)
          \right]^{r_\mathrm{o}}_{x''_\mathrm{op}}\\
       &=  (x''_\mathrm{op}-d_\mathrm{op})\sqrt{r_\mathrm{p}^{2}-(x''_\mathrm{op}-d_\mathrm{op})^{2}}
          +r_\mathrm{p}^{2}\sin^{-1}\left(\dfrac{x''_\mathrm{op}-d_\mathrm{op}}{r_\mathrm{p}}\right)\\
       &\;\;\;\; +r_\mathrm{p}^{2}\dfrac{\pi}{2}+r_\mathrm{o}^{2}\dfrac{\pi}{2}-x''_\mathrm{op}\sqrt{r_\mathrm{o}^{2}-
        (x''_\mathrm{op})^{2}}-r_\mathrm{o}^{2}\sin^{-1} \left(\dfrac{x''_\mathrm{op}}{r_\mathrm{o}}\right).
\end{aligned}
\end{equation*}

Similarly, changing $d_\mathrm{op}$, $r_\mathrm{o}$, ${x_\mathrm{op}''}$ in $k_\mathrm{op}(t)$ to $d_\mathrm{ep}$, $r_\mathrm{e}$, ${x_\mathrm{ep}'}$ respectively will provide $k_\mathrm{ep}(t)$: 

\begin{equation*}
\begin{aligned}
k_\mathrm{ep}(t) &= (x'_\mathrm{ep}-d_\mathrm{ep})\sqrt{r_\mathrm{p}^{2}-(x'_\mathrm{ep}-d_\mathrm{ep})^{2}}
          +r_\mathrm{p}^{2}\sin^{-1}\left(\dfrac{x'_\mathrm{ep}-d_\mathrm{ep}}{r_\mathrm{p}}\right)\\
       &\;\;\;\; +r_\mathrm{p}^{2}\dfrac{\pi}{2}+r_\mathrm{e}^{2}\dfrac{\pi}{2}-x'_\mathrm{ep}\sqrt{r_\mathrm{e}^{2}-
        (x'_\mathrm{ep})^{2}}-r_\mathrm{e}^{2}\sin^{-1} \left(\dfrac{x'_\mathrm{ep}}{r_\mathrm{e}}\right).
\end{aligned}
\end{equation*}

Define $k_\mathrm{eo}$ as the overlapping area of the active satellite and the umbra divided by the passive satellite circle area. Similar to $k_\mathrm{ep}(t)$ and $k_\mathrm{op}(t)$, changing $d_\mathrm{ep}$, $r_\mathrm{p}$, ${x_\mathrm{ep}'}$ in $k_\mathrm{ep}$ to $d_\mathrm{eo}$, $r_\mathrm{o}$, $x_\mathrm{eo}$ respectively will provide $k_\mathrm{eo}(t)$:

\begin{equation*}
\begin{aligned}
k_\mathrm{eo}(t) &= (x_\mathrm{eo}-d_\mathrm{eo})\sqrt{r_\mathrm{o}^{2}-(x_\mathrm{eo}-d_\mathrm{eo})^{2}}
          +r_\mathrm{o}^{2}\sin^{-1}\left(\dfrac{x_\mathrm{eo}-d_\mathrm{eo}}{r_\mathrm{o}}\right)\\
          &\;\;\;\; +r_\mathrm{o}^{2}\dfrac{\pi}{2}+r_\mathrm{e}^{2}\dfrac{\pi}{2}-x_\mathrm{eo}\sqrt{r_\mathrm{e}^{2}-
          {x}_\mathrm{eo}^{2}}-r_\mathrm{e}^{2}\sin^{-1} \left(\dfrac{x_\mathrm{eo}}{r_\mathrm{e}}\right).
\end{aligned}
\end{equation*}

\section*{Appendix C: Formulating $k(t)$ \label{app:k}}

$k(t)$ is the overlapping area of the passive satellite, the active satellite and the umbra at time $t$. Since $k(t)$ depends on the orientations of three circles, $k(t)$ is separated into different cases by considering the circle overlapping. 

\begin{table}[ht]
\centering
\caption{Graphical representations for finding $k(t)$. \label{tab:k_cases_dop}}
\centering
\begin{tabular}{|c|c|}
\hline
$d_\mathrm{eo} > r_\mathrm{o} + r_\mathrm{p}$  & $d_\mathrm{eo} < |r_\mathrm{o} - r_\mathrm{p}|$  \\
\hline
\parbox[c]{1in}{\centering \includegraphics[width=1in]{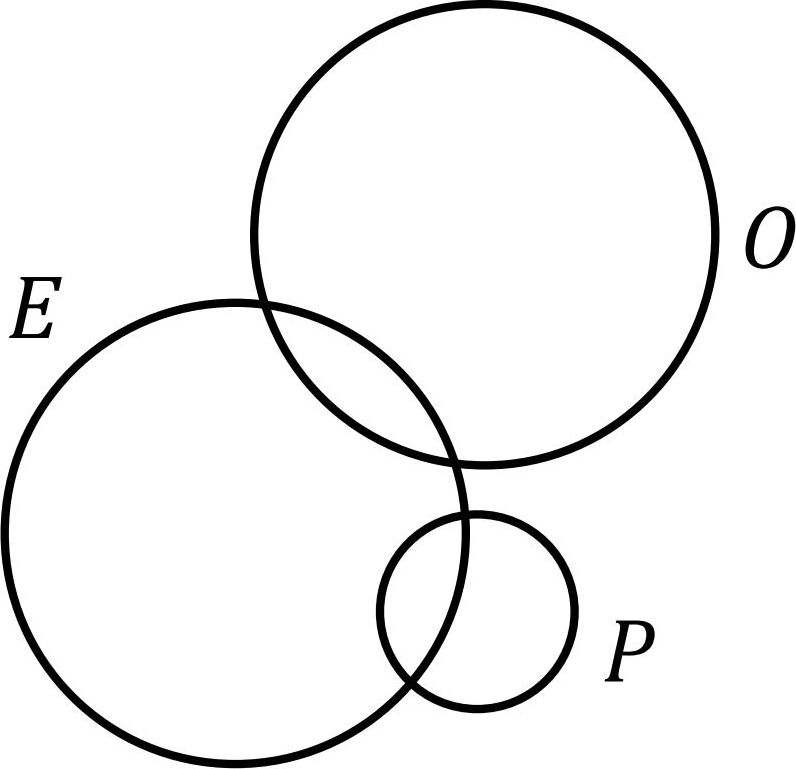}}  &  \parbox[c]{1in}{\centering \includegraphics[width=0.9in]{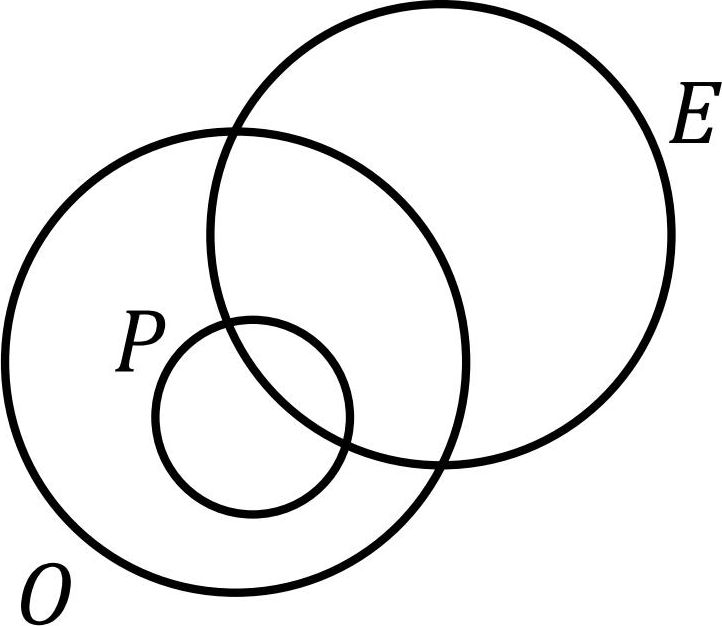}}  \\
\hline
\end{tabular}
\end{table}

First, if any two circles do not intercept with each other, $k(t)$ can be found easily. For example, if the circles $O$ and $P$ do not intercept with each other, either $d_\mathrm{eo} > r_\mathrm{o} + r_\mathrm{p}$ or $d_\mathrm{eo} < |r_\mathrm{o} - r_\mathrm{p}|$ (see Table~\ref{tab:k_cases_dop}). In the former case, $k(t)=0$. In the latter case, $k(t)$ equals to the overlapping area between the circle $E$ and the smaller one among the circles $O$ and $P$. Then, the configuration of three circles can be divided into different cases by three statements.
\begin{enumerate}
\item The interceptions between the circles $O$ and $E$ are both inside ($I$) or both outside ($O$) the circle $P$.
\item The interceptions between the circles $O$ and $P$ are both inside or both outside the circle $E$.
\item The interceptions between the circles $E$ and $P$ are both inside or both outside the circle $O$.
\end{enumerate}

For example, the case $IOI$ means that 
the circles $O$ and $E$ both intercept inside the circle $P$, 
the circles $O$ and $P$ both intercept outside the circle $E$, while
the circles $E$ and $P$ both intercept inside the circle $O$.
Note that the case $III$ does not exist. 

$k(t)$ of all cases can be determined by considering four points, $(x_\mathrm{eo},y_\mathrm{eo})$, $(x_\mathrm{eo},-y_\mathrm{eo})$, $(x_\mathrm{op},y_\mathrm{op})$, $(x_\mathrm{ep},-y_\mathrm{ep})$ and their distances from the center of the circles. See Table~\ref{tab:k_cases} for a summary. 

An additional case is that the interceptions are one inside and one outside of the third circle of all three circles. In that case, a circular triangle formed. So, to determine the circular triangle ($A_\mathrm{CT}$) case, both interceptions between E and O are checked, which have the coordinates $(x_\mathrm{eo},y_\mathrm{eo})$ and  $(x_\mathrm{eo},-y_\mathrm{eo})$. Appendix D introduces the calculation of $A_\mathrm{CT}$. 

\begin{table}[ht]
\centering
\caption{Cases of $k(t)$ and configurations of the circles $O$, $E$ and $P$.
\label{tab:k_cases}}
\centering
\begin{tabular}{|c|c|c|}
\hline
\textbf{Case} & \textbf{Configuration} & \textbf{$k(t)=$} \\
\hline
\rule{0pt}{10pt}$IIO$ & \multirow{3}{*}{\parbox[c]{1in}{\centering \includegraphics[width=0.6in]{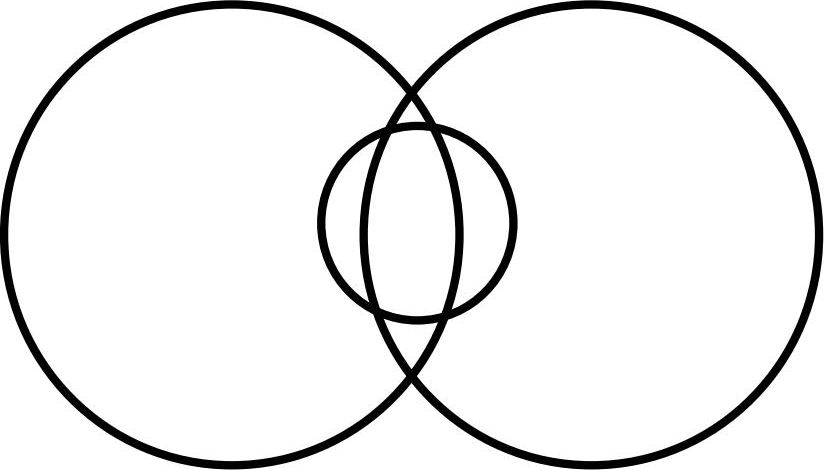}}} & $k_\mathrm{eo} + k_\mathrm{op} - \pi r_\mathrm{o}^{2}$      \\ 
$IOI$ &                  & $k_\mathrm{eo} + k_\mathrm{ep} - \pi r_\mathrm{e}^{2}$                \\  
$OII$  &                  & $k_\mathrm{op} + k_\mathrm{ep} - \pi r_\mathrm{p}^{2}$               \\  
\hline
\rule{0pt}{10pt}$IOO$ & \multirow{3}{*}{\parbox[c]{1in}{\centering \includegraphics[width=0.5in]{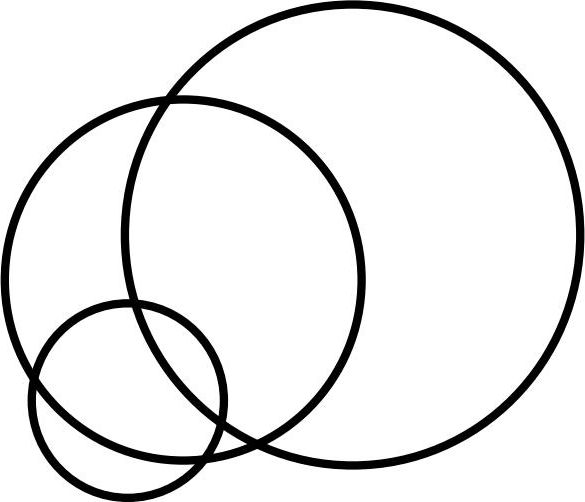}}} & $k_\mathrm{eo}$            \\ 
$OIO$  &                  & $k_\mathrm{op}$               \\ 
$OOI$  &                  & $k_\mathrm{ep}$                \\ 
\hline
$OOO$  &\parbox[c]{1in}{\centering \includegraphics[width=0.58in]{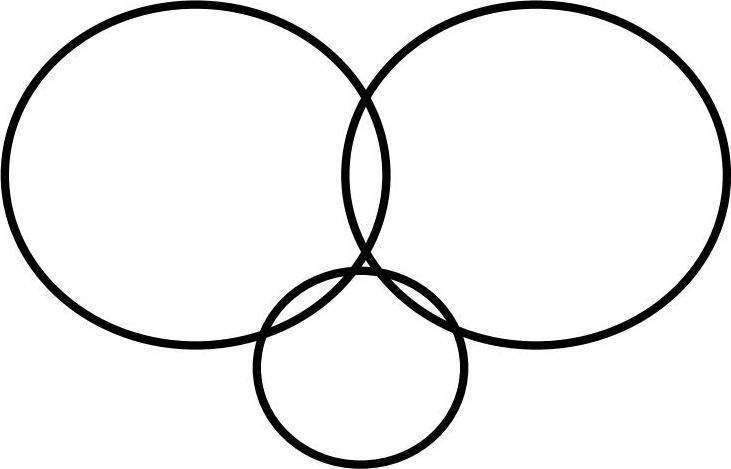}} & 0             \\
\hline
Circular triangle  & \parbox[c]{1in}{\centering \includegraphics[width=0.58in]{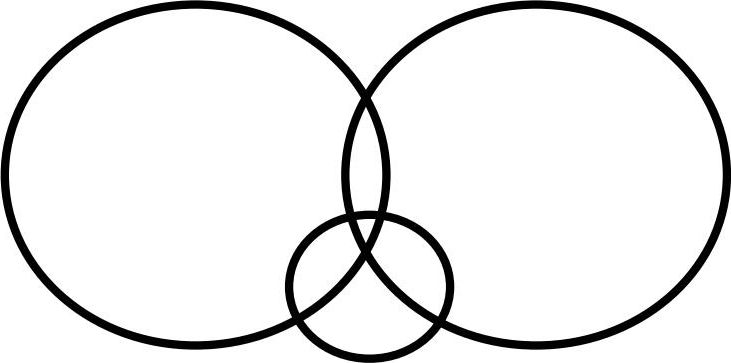}} & $A_\mathrm{CT}$  \\
\hline
\end{tabular}
\end{table}

\section*{Appendix D: Formulating $A_{\rm CT}$ \label{app:A}}

To find $A_\mathrm{CT}$ mentioned near the end of Appendix C, we need to find the lengths of the chords $c_\mathrm{e}$, $c_\mathrm{o}$ and $c_\mathrm{p}$ as illustrated in Fig.~\ref{fig:circular_triangle} first:

\begin{figure}[ht] 
\centering
\includegraphics[width=9 cm]{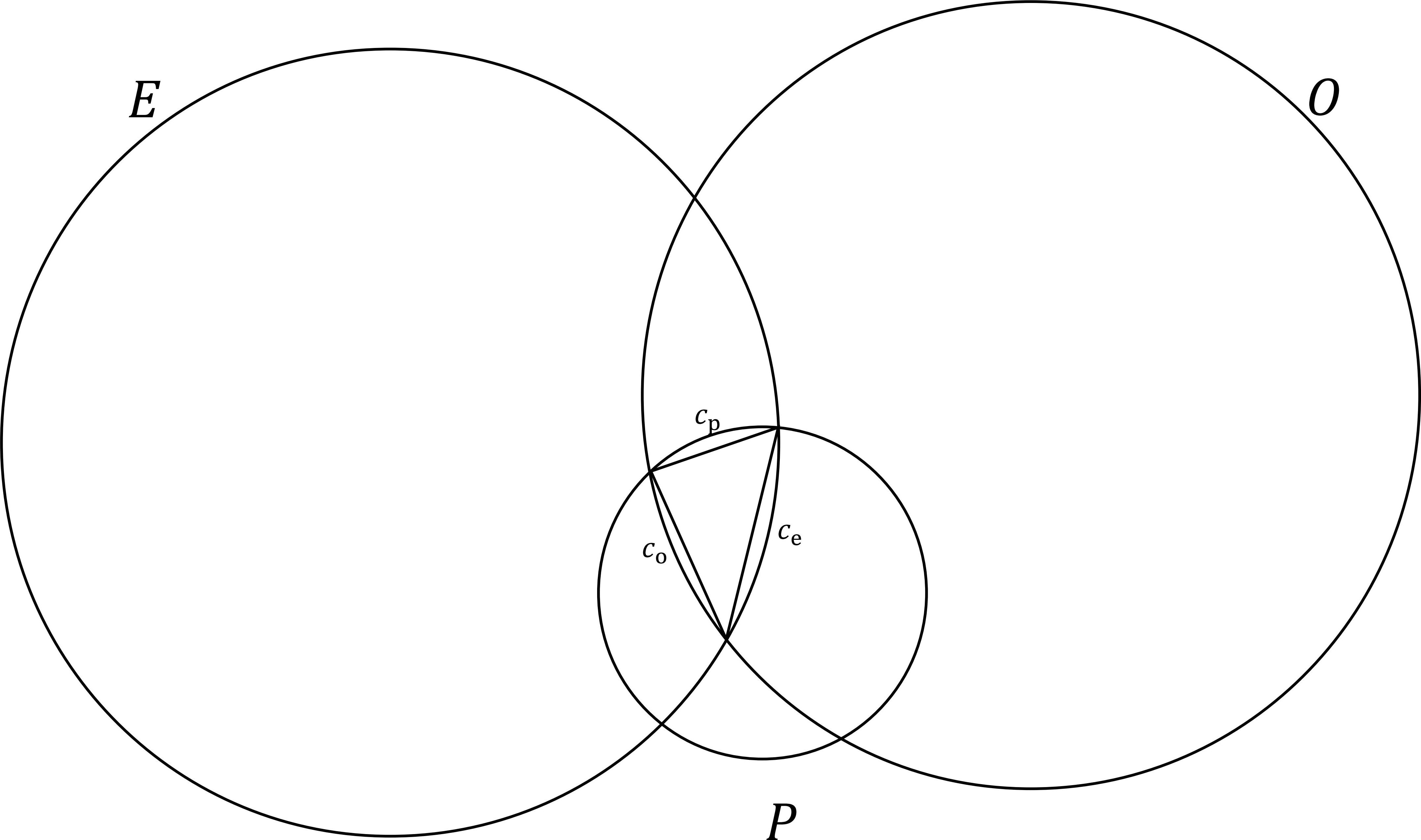}
\caption{Circular triangle and the triangle area formed by three chords and three segments. \label{fig:circular_triangle}}
\end{figure}

\begin{equation*}
\begin{aligned}
c_\mathrm{e} &= \sqrt{(x_\mathrm{ep}-x_\mathrm{eo})^{2}+(y_\mathrm{ep}-y_\mathrm{eo})^{2}} \mbox{\; ,}\\
c_\mathrm{o} &= \sqrt{(x_\mathrm{op}-x_\mathrm{eo})^{2}+(y_\mathrm{op}-y_\mathrm{eo})^{2}} \mbox{\; and}\\
c_\mathrm{p} &= \sqrt{(x_\mathrm{op}-x_\mathrm{ep})^{2}+(y_\mathrm{op}-y_\mathrm{ep})^{2}} \mbox{\; .}
\end{aligned}
\end{equation*}

Then $A_\mathrm{CT}$ is given by the area of the triangle formed by three chords and three segments:

\begin{equation*}
\begin{split}
A_\mathrm{CT} =& \frac{1}{4}\sqrt{(c_\mathrm{e}+c_\mathrm{o}+c_\mathrm{p})(c_\mathrm{e}+c_\mathrm{o}-c_\mathrm{p})(c_\mathrm{e}+c_\mathrm{p}-c_\mathrm{o})(c_\mathrm{o}+c_\mathrm{p}-c_\mathrm{e})}\\
   &+\left(r_\mathrm{o}^{2}\sin^{-1}\frac{c_\mathrm{o}}{2r_\mathrm{o}}-\frac{c_\mathrm{o}}{4}\sqrt{4r_\mathrm{o}^{2}-c_\mathrm{o}^{2}}\right)\\
   &+\left(r_\mathrm{e}^{2}\sin^{-1}\frac{c_\mathrm{e}}{2r_\mathrm{e}}-\frac{c_\mathrm{e}}{4}\sqrt{4r_\mathrm{e}^{2}-c_\mathrm{e}^{2}}\right)\\
   &+\left(r_\mathrm{p}^{2}\sin^{-1}\frac{c_\mathrm{p}}{2r_\mathrm{p}}-\frac{c_\mathrm{p}}{4}\sqrt{4r_\mathrm{p}^{2}-c_\mathrm{p}^{2}}\right)\mbox{\; .}
\end{split}
\end{equation*}

However, if more than half of the circle $P$ is included in the circular triangle, i.e., the inequality
\begin{equation*}
d_\mathrm{ep}\sin\theta'<y_\mathrm{ep}+\dfrac{y_\mathrm{op}-y_\mathrm{ep}}{x_\mathrm{op}-x_\mathrm{ep}}(d_\mathrm{ep}\cos\theta'-x_\mathrm{ep})
\end{equation*}
is true, then the last term of $A_\mathrm{CT}$ becomes $r_\mathrm{p}^{2}\left(\pi-\sin^{-1}\frac{c_\mathrm{p}}{2r_\mathrm{p}}\right)+\frac{c_\mathrm{p}}{4}
\sqrt{4r_\mathrm{p}^{2}-c_\mathrm{p}^{2}}$ \mbox{\; .}

\section*{Appendix E: Astrometric discrepancies originating from the projection planes mismatch\label{app:phase_angle}}
Previous studies had two different choices of coordinate projection planes to describe mutual eclipses and mutual occultations separately. 
\textit{MULTI-SAT} predicts the satellite's longitude relative to the projection of the vector of the \textit{Sun} on the equator of the planet (planetocentric planet-equatorial coordinate plane) for mutual eclipses. 
On the other hand, \textit{MULTI-SAT} predicts the satellite's longitude relative to the projection of the vector of the \textit{Earth} on the equator of the planet (synodic planetocentric coordinate plane) for mutual occultations \citep{emelyanov(2020b)}. The origin is placed at the eclipsed or occulted satellite.

\begin{figure}[ht] 
\centering
\includegraphics[width=9 cm]{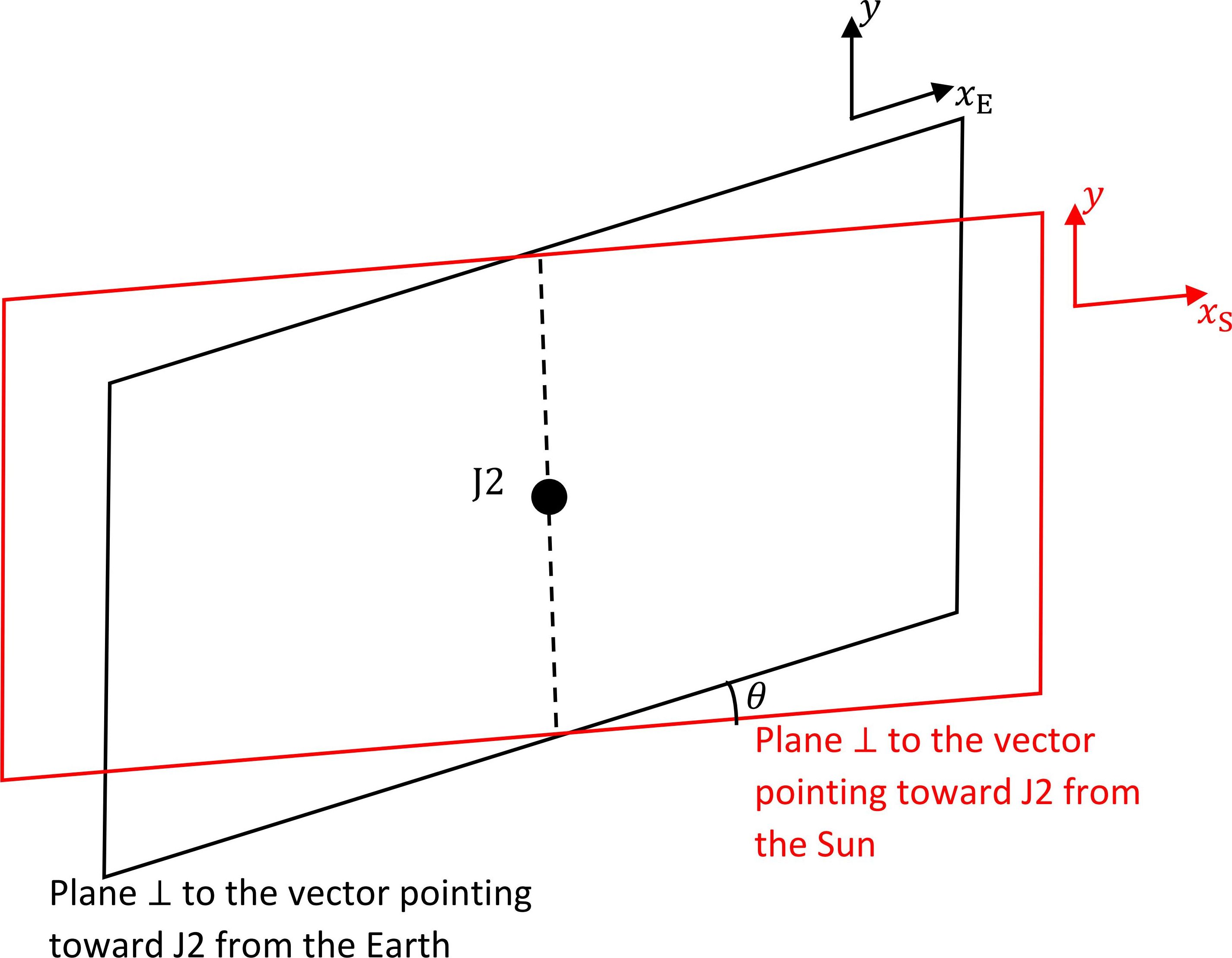}
\caption{Two different choices of coordinate projection planes to describe mutual eclipses and mutual occultations. \label{fig:coordinate_plane}}
\end{figure}

While the above choices of coordinate planes are justified for individual QEs, we have to choose one of them if QSME is considered. Our model in Sect.~\ref{sec:model} adopted the synodic coordinate plane in which the projection vector points from the observer on the Earth to the satellites. Since our choice favours mutual occultations, one may expect certain astrometric discrepancies when describing mutual eclipses. The discrepancies can be significant if the satellite's phase angle, i.e., the angle between coordinate planes, is larger (See Fig.~\ref{fig:coordinate_plane}). 

\begin{figure}[ht] 
\centering
\includegraphics[width=4 cm]{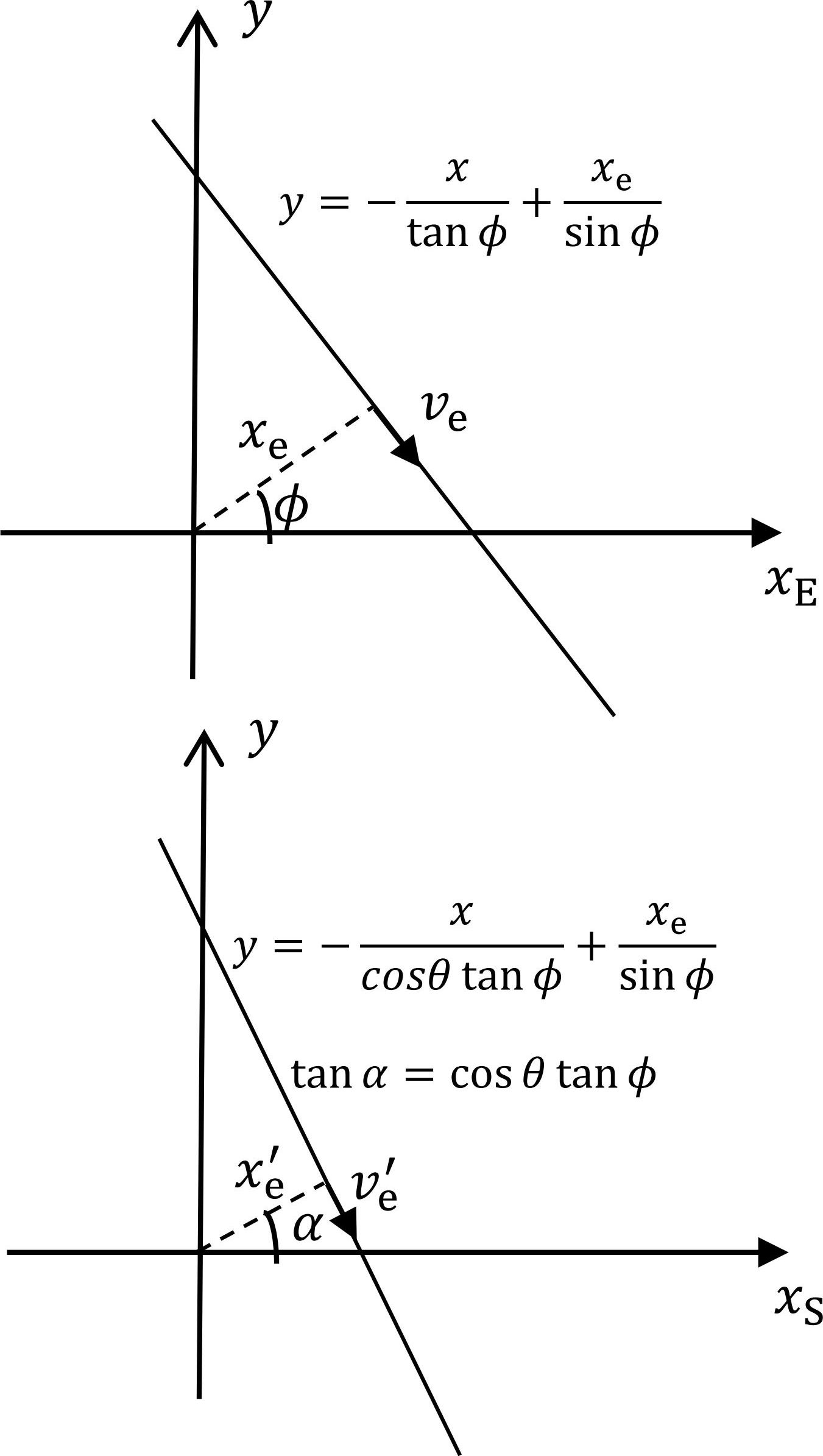}
\caption{Coordinates of the satellite's locus on different coordinate projection planes. \label{fig:locus}}
\end{figure}

From NASA JPL's \textit{Horizons System}\footnote{\url{https://ssd.jpl.nasa.gov/horizons/app.html}}, the phase angle during the QSME ranged from $0.680^{\circ}$ to $0.695^{\circ}$. To estimate the size of the discrepancies, let us consider the $x$ coordinates of the satellite's locus on both planes. From Fig.~\ref{fig:locus}, the locus of J3 on the synodic coordinate plane is given by
\begin{equation}
    \begin{cases}
    x_\mathrm{E} = v_\mathrm{e}(t-t_\mathrm{e})\sin\phi+x_\mathrm{e}\cos\phi\\
    y = -v_\mathrm{e}(t-t_\mathrm{e})\cos\phi+x_\mathrm{e}\sin\phi\\
    \end{cases} 
    \mbox{\; .}
\label{eqt:locus_e_synodic}
\end{equation}
for an unknown value of $\phi$ in the $(y,x_\mathrm{E})$ coordinate. 

On the planet-equatorial coordinate plane $(y,x_\mathrm{S})$, since $x_\mathrm{E}\cos\theta = x_\mathrm{S}$, the locus is given by
\begin{equation}
    \begin{cases}
    x_\mathrm{E} = v_\mathrm{e}(t-t_\mathrm{e})\sin\phi\cos\theta+x_\mathrm{e}\cos\theta\cos\phi\\
    y = -v_\mathrm{e}(t-t_\mathrm{e})\cos\phi+x_\mathrm{e}\sin\phi\\
    \end{cases}     
    \mbox{\; .}
\label{eqt:locus_e_equatorial}
\end{equation}

To find $x_\mathrm{e}'$, we use the fact that there are two ways to calculate the area of the right angle triangle,
\begin{equation}
\frac{1}{2}x_\mathrm{e}'\sqrt{\left(\frac{x_\mathrm{e}}{\sin\phi}\right)^2+\left(\frac{x_\mathrm{e}\cos\theta}{\cos\phi}\right)^2} = \frac{1}{2}\left| \frac{x_\mathrm{e}}{\sin\phi}\right|\left| \frac{x_\mathrm{e}\cos\theta}{\cos\phi}\right|
    \mbox{\; ,}
\end{equation}
\begin{equation}
x_\mathrm{e}' = \frac{\left|\cos\theta\right|}{\sqrt{\sin^2\phi\cos^2\theta+\cos^2\phi}}x_\mathrm{e}
    \mbox{\; .}
\end{equation}
The maximum change in $x_\mathrm{e}$ occurred when $\phi=0$ or $\phi=\pi$. In these cases, $x_\mathrm{e}'=x_\mathrm{e}\cos\theta$.

To find $v_\mathrm{e}'$,
\begin{equation}
v_\mathrm{e}'= \sqrt{(\dot{x_\mathrm{E}})^2+(\dot{y})^2} = v_\mathrm{e}\sqrt{\sin^2\phi\cos^2\theta+\cos^2\phi}
    \mbox{\; .}
\end{equation}
So the maximum change in $v_\mathrm{e}$ occurred when $\phi=\pi/2$ or $\phi=3\pi/2$. In these cases, $v_\mathrm{e}'=v_\mathrm{e}\cos\theta$. 

To find $t_\mathrm{e}'$,
\begin{equation}
x_\mathrm{e}'\cos\alpha = v_\mathrm{e}(t_\mathrm{e}'-t_\mathrm{e})\sin\phi\cos\theta+x_\mathrm{e}\cos\theta\cos\phi
    \mbox{\; ,}
\end{equation}
\begin{equation}
\frac{\cos\theta}{\sqrt{\sin^2\phi\cos^2\theta+\cos^2\phi}}\cos\alpha = v_\mathrm{e}(t_\mathrm{e}'-t_\mathrm{e})\sin\phi\cos\theta + x_\mathrm{e}\cos\theta\cos\phi
    \mbox{\; .}
\end{equation}
Since $\left|\cos\alpha\right| = {1}/{\sqrt{\tan^2\alpha+1}} = {1}/{\sqrt{\cos^2\theta\tan^2\phi+1}}$, 
\begin{equation}
\left|t_\mathrm{e}'-t_\mathrm{e}\right|=\frac{x_\mathrm{e}}{v_\mathrm{e}}\left|\frac{\cos\phi\sin\phi\sin^2\theta}{\sin^2\phi\cos^2\theta+\cos^2\phi}\right|
    \mbox{\; .}
\end{equation}

Plugging in our fitted values from Table~\ref{tab:results}, the astrometric discrepancies on $x_\mathrm{e}$ (in arcsec), $v_\mathrm{e}$ (in arcsec hr$^{-1}$) and $t_\mathrm{e}$ (in hr) due to the projection planes mismatch are limited to
$(1-\cos0.695^\circ) \times x_\mathrm{e} = 7.6\times10^{-6}$,
$(1-\cos0.695^\circ) \times v_\mathrm{e} = 2.1\times10^{-4}$ and 
$0.00614\times{x_\mathrm{e}}/{v_\mathrm{e}} = 2.2\times10^{-4}$
respectively. They are at least one order of magnitude smaller than the parameter errors. To conclude, in view of the smallness of the phase angle, we can safely ignore the astrometric discrepancies and compute their O-Cs directly.  
\end{document}